\documentclass[twocolumn,amsmath,amssymb,eqsecnum,nofootinbib,aps,prd,10pt]{revtex4-2}

\usepackage{graphicx}
\graphicspath{{./images/}}
\usepackage{mathtools}
\usepackage[scr=esstix]{mathalfa}
\usepackage{hyperref}
\usepackage[noabbrev]{cleveref}

\usepackage{xcolor}
\usepackage{cancel}
\usepackage[normalem]{ulem}

\makeatletter
\def\l@subsubsection#1#2{}
\makeatother
\crefname{equation}{}{}

\newcommand{\realn}{\mathbb{R}}

\DeclareMathOperator{\Tr}{Tr}

\newcommand{\ph}{\varphi}
\newcommand{\tht}{\vartheta}
\newcommand{\eps}{\varepsilon}
\newcommand{\kap}{\varkappa}
\newcommand{\Chi}{{\mathcal{X}}}

\newcommand{\ts}[1]{{\boldsymbol{#1}}}
\newcommand{\dif}{\ts{d}}

\newcommand{\Ric}{\ts{\mathrm{Ric}}}
\newcommand{\Ein}{\ts{\mathrm{Ein}}}
\newcommand{\scR}{\mathcal{R}}
\newcommand{\spsc}{{\mathscr{r}}}

\newcommand{\Terg}{\ts{T}}
\newcommand{\eTS}{{\sigma_\Sigma}}
\newcommand{\pTS}{{\tau_\Sigma}}
\newcommand{\ehTS}{{\hat\sigma}}
\newcommand{\phTS}{{\hat\tau}}

\newcommand{\mBTZ}{\hat{M}}
\newcommand{\MKV}{\mathcal{M}}
\newcommand{\locmass}{\mathscr{m}}

\newcommand{\AdS}{{\mathrm{AdS}}}
\newcommand{\BTZ}{{\mathrm{BTZ}}}

\newcommand{\ai}{{\mathrm{a}}}
\newcommand{\oi}{{\mathrm{o}}}
\newcommand{\iix}{{\mathrm{in}}}
\newcommand{\oix}{{\mathrm{out}}}
\newcommand{\hor}{{\mathrm{hor}}}

\newcommand{\fld}{{\mathrm{fluid}}}
\newcommand{\dst}{{\mathrm{dust}}}
\newcommand{\shl}{{\mathrm{shell}}}

\newcommand{\dom}{\mathcal}
\newcommand{\const}{\text{const}}
\newcommand{\xx}{x}

\begin{document}

    \title{Models of a point particle in a three-dimensional AdS universe}
    
    \author{Petr Luke\v{s}}
    \email{petr.lukes934@student.cuni.cz}
    \author{Pavel Krtou\v{s}}%
    \email{pavel.krtous@utf.mff.cuni.cz}
    \affiliation{%
        Institute of Theoretical Physics,\\
        Faculty of Mathematics and Physics, Charles University\\
        V Hole\v{s}ovi\v{c}k\'{a}ch 2, 180 00 Prague 8, Czech Republic
    }%
    
    \date{\today}

    \begin{abstract}\noindent
        In 2+1-dimensional gravity, a conical deficit is commonly interpreted as a point-like particle. A distributional description of such a source is problematic since the Einstein equations are non-linear. Moreover, the total mass of the spacetime with nontrivial asymptotics is obfuscated by the infinite amount of background energy in the distant regions. In this work, we support the standard claim that the mass of a point particle is given by the angular deficit of the asymptotic geometry. We approximate the singular matter source by a sufficiently regular energy distribution, and we study the limit of negligible size for such a distribution while keeping the asymptotic geometry intact. Comparing spacetimes with equal asymptotics ensures that we correctly identify and remove the cosmological matter. We find that the angular deficit of the resulting conical spacetime is given by a limit of the local mass of the object, where the local mass is a natural concept of additive mass in 2+1 gravity. Surprisingly, the limiting mass is different from the BTZ mass parameter, although both masses are closely related.
    \end{abstract}
    
    \maketitle

    \section{Introduction}

The main goal of this work is to better understand the mass of a singular point-like matter source in \mbox{2+1-dimensional} General Relativity. A point-like particle in three spacetime dimensions is usually described by a conical singularity \cite{Levi-Civita:2011,Marder:1959,Staruszkiewicz:1963}, see also \cite{Vilenkin:1981,Gott:1984}. While such a particle is often understood as a $\delta$-distribution of mass \cite{Deser:1984a,Deser:1984b,Deser:1985}, in nonlinear theories such as General Relativity, this description may not be well-founded, since distributional theory is mostly suited to linear theories. One should employ a more general mathematical description, such as \cite{Grosseretal:2014}, or try to approximate singular sources using a suitably regular matter distribution. 

Despite these problems, there is general agreement that the mass of the point particle in 2+1 dimensions is related to the deficit angle of the conical singularity. We want to support this fact by presenting various sufficiently regular matter models of compact objects which, in the limit of negligible size, approach the conical singularity. We demonstrate that a natural notion of local mass of the object yields the standard relation between the mass of a point-like object and the conicity.

While a point-like distribution of mass is mathematically problematic, it is well known that a shell-like object with the worldsheet of spacetime codimension one can be described by a careful treatment of the junction conditions for the geometry \cite{Israel:1966,Barrabes:1991}. In three spacetime dimensions, such objects describe both domain-wall-like branes and string-like objects simultaneously. Recently, in \cite{Lukes:2026a}, we studied accelerated point-like sources that interact with strings, generalizing the C-metric solution in 2+1 gravity \cite{Astorino:2011}. In this paper, we will use a thin shell located on the surface of a spherical matter object. Although these objects are also non-regular, they do not cause mathematical problems such as point-like singularities do.  

The notion of mass in a curved spacetime is notoriously hard to define unambiguously. The local mass density is the time-time component of the stress energy tensor \cite{Einstein:1922}. A description of the total mass of a spacetime is generally much more complex \cite{Arrowit:1959}, especially if it intends to include a contribution of gravitational energy. Fortunately, the situation is simpler in 2+1 dimensions, where gravity has no local degrees of freedom, and one does not expect its contribution to the total mass count. However, the local mass density depends on the choice of the time slices. There can be a preferred slicing in spacetime, often based on a time-like symmetry, which gives the mass a better meaning \cite{Komar:1959}. In this work, we will focus on static spacetimes with a time-like symmetry. It allows us to define two plausible concepts of the total mass: (i) a plain integral of the local mass density over a static slice and (ii) the Killing mass related to the static Killing vector, cf.~\cite{Tolman:1930}. It turns out that the first concept yields the expected relationship between particle mass and conicity. 

Another issue arises when a spacetime is spatially infinite: a non-vanishing stress-energy distribution (including a dark-energy contribution related to a non-trivial cosmological constant) might be present and cannot be localized within a finite subregion. In such a case, the integral for the local mass provides an infinite result, which obfuscates the meaning of such an integral. It can be resolved in two ways. Either by focusing on asymptotically flat spacetimes only or by identifying some standard infinite contribution to the mass (such as the contribution of dark energy) and ``renormalizing'' it away, e.g., as in \cite{Abbot:1982,Magnon:1985,Brown:1993}. 

In our approach, we consider matter objects localized in a compact region, and we do not include the dark-energy contribution outside the object.  When we consider spacetimes with matter objects of various sizes, we compare only spacetimes with the same asymptotic behavior. Thus, we can claim that we have removed the same asymptotic infinite amount of mass related to dark energy distributed throughout the whole asymptotic region.

In four dimensions, where the problem is more complicated, many other attempts have been made to circumvent the problems mentioned above by introducing notions collectively called quasi-local masses. Their definitions vary greatly (e.g., \cite{Tolman:1930,Miesner:1964,Hawking:1968,Geroch:1973,Bartnik:1989,Penrose:1982} and many others).

In \Cref{sec:statrotsym} we set up the basic mathematical framework for describing static, rotationally symmetric spacetimes. In \cref{sec:AsymptGeom}, we discuss $\Lambda$-vacuum spacetimes that may possess a central singularity. In \Cref{sec:matter}, we introduce matter sources used later in constructing models of compact objects. In \Cref{sec:mass} we define the above-mentioned notions of mass localized in a spatial domain. In \Cref{sec:combine}, we construct several sufficiently regular models of compact matter objects that can approximate the point-like singular source. We then study the limit when the size of such objects approaches zero. In particular, we study the behavior of the mass of the object in that limit. We summarize the results in \cref{sec:summary}.
    \section{Static-rotation symmetric solutions of Einstein equations in 3D}
\label{sec:statrotsym}

This work focuses on solutions of the unmodified Einstein equations
\begin{equation}
    \label{eq:ein_eq}
    \Ein + \Lambda\ts{g} = \kap\,\Terg
\end{equation}
in 2+1 dimensions. $\Ein = \Ric - \frac{1}{2}\ts{g}\scR$ is the Einstein tensor, and $\Terg$ is the stress-energy tensor describing matter. $\Lambda$ is the cosmological constant, which may also be interpreted as dark energy -- a uniform distribution of density and pressure in spacetime.

In 2+1 dimensions, the prominent feature of the Riemann tensor is that it has the same number of degrees of freedom as its trace, the Ricci tensor \cite{Fock:1962}. Since the Ricci tensor is fully determined by \cref{eq:ein_eq}, so is the Riemann tensor. Consequently, spacetime is curved only where matter is located (including dark matter). There are no other gravitational degrees of freedom, no gravitational waves, nor even attraction between massive bodies. Yet, interesting features can be exhibited in the global geometry, often related to spacetime defects\footnote{While the defects could perhaps be included in \cref{eq:ein_eq}, their singular character would require a distributional description likely extended beyond usual distributional theory, and therefore we treat them here separately on a more intuitive basis.} and topological changes\cite{Staruszkiewicz:1963,Deser:1984a,Deser:1984b,Deser:1989}. 

In this work, we assume the static and rotationally symmetric ansatz, allowing the spacetimes to include singular matter distributions as a point particle at the origin and spherical shells. Therefore, one must be careful when calculating integral observables containing these singular contributions. 

Some of the discussion is standard material found in textbooks, especially in the four-dimensional context \cite{Misner:1973}. However, the details of the limiting procedures discussed below may clarify subtleties of the singular distributions and nature of mass parameters.

Static symmetry implies the existence of a time-like, hypersurface orthogonal Killing vector with orbits of topology $\realn$. Rotational symmetry implies a space-like Killing vector with compact orbits of topology $S^1$. Both Killing vectors commute. 

Our assumptions allow us to choose Killing coordinates $T\in\realn$ and $\ph\in({-\pi},{\pi})$ associated with the static and rotational Killing vectors $\ts{\partial}_T$ and $\ts{\partial}_\ph$, respectively, identifying the surfaces $\ph = \pm\pi$. The spacetime metric can be written as
\begin{equation}
    \label{eq:statrotmetricTrph}
	\ts{g} = - N^2 \dif T^2 + \dif r^2 + R^2 \dif\ph^2\;,
\end{equation}
with metric functions $N=N(r)$ and $R=R(r)$ depending only on the radial coordinate $r$. Clearly, $r$ measures the radial proper distance, and we naturally require $R\ge0,\ N\ge0$.

The geometry \eqref{eq:statrotmetricTrph} can also be rewritten using the circumference-radial coordinate $R$
\begin{equation}
    \label{eq:statrotmetricTRph}
	\ts{g} = - N^2 \dif T^2 + \frac{\dif R^2}{R'{}^2}\, + R^2 \dif\ph^2\;,
\end{equation}
where we now understand the metric functions as functions of $R$, $N=N(r(R))$, and $R'=R'(r(R))$, the prime representing differentiation with respect to $r$. 

In the smooth case, both descriptions are equivalent. However, if we also allow non-smooth spacetimes containing a point source or shells, the smooth differential structure is given by coordinates\footnote{%
Of course, polar coordinates $T,r,\ph$ alone do not define a differential atlas. One has to use more versions of them to cover the semi-axis $\ph=\pm\pi$. The origin must also be covered by other coordinates. Unfortunately, for spacetime with a conical deficit, the differential and tangent structure is broken at the origin. But we try to avoid a technical discussion of this issue and instead present arguments suggesting how to approach the conical deficit as a limit of smooth spacetimes.\\
Another issue arises when describing a spherical shell with singular behavior at $r=r_*$. The coordinate $r$ describes a smooth structure across the shell. We have the freedom to shift such a coordinate $r$ by a constant. However, it will be advantageous to make a different shift on both sides of the shell, thus making $r$ non-regular at the shell. But the relation to the smooth variant of $r$ is simple and obvious.}
$T,r,\ph$. In contrast, the metric function $R(r)$ need not be smooth. It should be continuous, but a discontinuity in $R'$ corresponds to the presence of a thin shell. Coordinates $T, R, \ph$ thus do not generally belong to the smooth atlas.

The Einstein tensor of spacetime geometry is 
\begin{equation}
    \label{eq:statrotEin}
	\Ein = - \frac{R''}{R}\,N^2 \dif T^2 + \frac{N'R'}{N R}\,\dif r^2 + \frac{N''}{N}\,R^2 \dif\ph^2\;,
\end{equation}
We assume the matter ansatz of a rotationally-symmetric fluid with a stress-energy tensor.
\begin{equation}
    \label{eq:statrotT}
	\Terg = \eps\,N^2 \dif T^2 + p_r\,\dif r^2 + p_\ph\,R^2 \dif\ph^2\;,
\end{equation}
with energy density $\eps$ and pressures $p_r$, $p_\ph$ depending only on $r$. For the fluid distributed in all spacetime dimensions, we always assume the isotropy of pressure, i.e.,
\begin{equation}
    \label{eq:pressureisotropy}
	  p\equiv p_r = p_\ph\;.
\end{equation}
In \cref{sec:matter}, we will discuss some simple equations of state that constrain the relationship between energy density and pressure.

We also introduce dimensionless variants of these quantities
\begin{equation}\label{eq:rescaledepsp}
     \hat\eps=\kap\eps\ell^2\;,\quad \hat{p}=\kap p\ell^2\;.
\end{equation}

The Einstein equations (now with the cosmological term implicitly included in the stress-energy tensor)
\begin{equation}
    \label{eq:Einsten}
	\Ein = \kap\, \Terg
\end{equation}
give
\begin{equation}
    \label{eq:fluidEq}
	\kap\eps = -\frac{R''}{R}\;,\quad 
	\kap p_r = \frac{N'R'}{N R}\;,\quad 
    \kap p_\ph = \frac{N''}{N}\;.
\end{equation}
The pressure-isotropy condition \cref{eq:pressureisotropy} implies a restriction on the metric functions:
\begin{equation}
    \label{eq:fluidisotropy}
    \frac{R'}{R}=\frac{N''}{N'} \quad\Rightarrow\quad 
    \frac{N'}{R}=\const\;.
\end{equation}

A two-dimensional spatial section ${T=\const}$ has a metric
\begin{equation}
    \label{eq:spatmetric}
	\ts{q} = \dif r^2 + R^2 \dif\ph^2 = \frac{\dif R^2}{R'{}^2}\, + R^2 \dif\ph^2 \;.
\end{equation}
Its Gauss curvature (half of the scalar curvature~$\spsc$) is given by the temporal component\footnote{The time-time component $\Ein_{\perp\perp}=  \ts{u} \cdot\Ein\cdot \ts{u}$ is calculated with respect to the normalized velocity $\ts{u}=\frac{1}{N}\ts{\partial}_T$ of the static observers.} of the spacetime Einstein tensor 
\begin{equation}
    \label{eq:EinSccur}
    \frac12\,\spsc =\Ein_{\perp\perp}=  -\frac{R''}{R}\;.
\end{equation}

This spatial rotation-symmetric geometry can have either no origin (``wormhole/bridge'' topology joining two asymptotic regions), one origin (``ordinary'' space with one asymptotic region), or two origins (``bubble'' topology). The length of an orbit of the rotational symmetry is $2\pi R$. Following its decreasing direction, one can approach zero, indicating the presence of the origin. We speak about a semi-regular origin if the length of the orbit is proportional to the proper distance from the origin. If the factor of proportionality is $2\pi$, the origin is regular; otherwise, there is a conical defect interpreted as a point particle. 

The lengths of the orbits can also decrease to a minimal non-zero value. Such an orbit is interpreted as the neck of a spatial wormhole. The most interesting case occurs when it corresponds to the black hole horizon. In such a case, the Killing vector of the time symmetry has a bifurcation character there. 

The presence of the origin at ${r = r_\oi}$ is equivalent to ${R(r_\oi)=0}$, the semi-regularity means $R'(r_\oi)\neq 0,\infty$, and the regularity of the origin requires $|R'(r_\oi)| = 1$.  In the wormhole-like case, $r_\oi$ corresponds to the smallest length orbit, and we require ${R'(r_\oi) = 0}$. The irregularity of $R'$, particularly ${R'(r_\oi)\neq 0}$, would imply the presence of a thin shell.  

For the lapse function, we denote the value at $r_\oi$ as ${N_\oi \equiv N(r_\oi)}$, and for a semi-regular origin, we assume ${N_\oi>0}$ and ${N'(r_\oi) = 0}$. The static wormhole with the neck at $r_\oi$ corresponds to ${N_\oi> 0}$; the horizon occurs for ${N_\oi = 0}$.

In general, the normalization $N_\oi$ can be changed by rescaling the Killing coordinate $T$. Special choices are related to the choice of normalization of the static Killing vector at particular places, in particular, at the origin or at infinity. Beware, however, that in the AdS context, the static Killing vector cannot be normalized to a finite value at infinity.

    \section{Asymptotic geometries}
\label{sec:AsymptGeom}

We start with the standard discussion of the asymptotics of static rotationally symmetric $\Lambda$-vacuum spacetimes. These spacetimes are discussed in the literature, e.g. \cite{Deser:1984a,Deser:1984b,Banados:1992a}, but we review them here to unify the notation.


\subsection{Flat case, $\Lambda=0$}
\label{ssec:AsymptMink}

For $\Lambda=0$, the Einstein equations \cref{eq:fluidEq} imply that $N$ and $R$ are linear functions of $r$, and at least one of $N'$ and $R'$ is vanishing. 

Let's start with the case $R'\neq 0$, $N'=0$,
\begin{equation}\label{eq:MinkConeGeom}
\begin{gathered}
    N=N_\ai\;,\quad R= c_\ai (r-r_\oi)\;,\\
    \ts{g} = - N_\ai^2 \dif T^2 + \dif r^2 + r^2 c_\ai^2\,\dif\ph^2\;.
\end{gathered}    
\end{equation}
We can always use the freedom in shifting $r\to r+r_\oi$ to set $r_\oi=0$. Rescaling the Killing coordinate, $t=N_\ai T$, also yields.
\begin{equation}
    \ts{g} = - \dif t^2 + \dif r^2 +  r^2 c_\ai^2\,\dif\ph^2 \;.
\end{equation}
Recall that $\ph\in(-\pi,\pi)$ is $2\pi$-periodic. This is the reason why the situation with the coefficient $c_\ai$ is slightly different. Introducing a rescaled Killing coordinate $\phi = c_\ai\ph$ eliminates the coefficient $c_\ai$ from the metric,
\begin{equation}\label{eq:Minkmtrcphi}
    \ts{g} = - \dif t^2 + \dif r^2 +  r^2 \dif\phi^2 \;.
\end{equation}
but the coordinate $\phi\in(-\phi_\oi,\phi_\oi)$ becomes $2\phi_\oi$-periodic where $\phi_\oi=c_\ai\pi$. Clearly, it is a locally flat metric written in polar coordinates. However, because of the $2\phi_\oi$ periodicity, the geometry is not regular at the origin $r=0$. It describes a cone with a vertex angle $2\pi c_\ai$. Indeed, it can be obtained by cutting the wedge from the flat plane along straight semi-lines $\phi=\pm\phi_\oi$ and gluing the edges together. 

For later reference, we rewrite the metric \cref{eq:MinkConeGeom} using a different parameter\footnote{The hat above mass or other quantities indicates a dimensionless version of the quantity. The full-dimensional mass $M$ would be rescaled by the gravitational constant $\kap$ and a customary numerical factor, namely $\mBTZ=\frac{\kap M}{2\pi}$. We use the dimensional version of other mass parameters below, but we employ $\mBTZ$ in the case of BTZ mass.} 
\begin{equation}\label{eq:BTZconicityL0}
    1-\mBTZ \equiv  c_\ai^2 = \frac{\phi_\oi^2}{\pi^2}\;.
\end{equation} 
and normalize $\ts{\partial}_T$ so that $N_\ai^2=(1-\mBTZ)$. We obtain a form resembling the BTZ metric (cf.~below),
\begin{equation}\label{eq:MinkConeBTZ}
    \ts{g} = - (1-\mBTZ)\, \dif T^2 + \frac{1}{1-\mBTZ}\dif R^2 +  R^2 \dif\ph^2 \;.
\end{equation}
In the following sections, we justify that the conical geometry can be interpreted as a point-like particle characterized by a BTZ mass parameter~$\mBTZ$. Another common characterization of the particle mass is 
\begin{equation}\label{eq:locmassconicity}
    \hat\locmass_\oi \equiv  1-c_\ai = \frac{\Delta\phi_\oi}{\pi}\;,
\end{equation}
where ${2\Delta\phi_\oi=2(\pi-\phi_\oi)}$ is the angular deficit of the conical singularity. Obviously, both mass parameters are related
\begin{equation}\label{eq:locmassBTZmass}
    1-\mBTZ = (1-\hat\locmass_\oi)^2\;.
\end{equation}

Let us turn now to the case $N'\neq 0$, which enforces $R$ to be constant $R=R_\oi$. The lapse $N=N_\oi + C_\oi r$ can be modified to $N=r$ by shifting $r\to r-\frac{N_\oi}{C_\oi}$ and introducing $\tau=C_\oi T$, 
\begin{equation}\label{eq:MinkRind}
    \ts{g} = - r^2 \dif \tau^2 + \dif r^2 + R_\oi^2\, \dif\ph^2\;.
\end{equation}
This is again a locally flat metric, now written in Rindler coordinates, however, with the $2\pi$-periodic coordinate $\ph$. It would be Minkowski spacetime for $\ph\in\realn$ in the frame of accelerated observers moving along orbits of $\ts{\partial}_\tau$. The compactification $\ph\in(-\pi,\pi)$ selects only compact circles in the direction transversal to the acceleration. The spatial section of the acceleration horizon $r=0$ is thus compact, and the causal structure of the resulting space resembles that of black holes. However, the length of the circle $r=\const$ is $2\pi R_\oi$ and remains the same for all $r$. The spacetime thus does not have a sufficiently growing asymptotic region for $r\to\infty$. Therefore, the spacetime is not usually considered physically interesting.

The final case is $N$ and $R$ constant, $N=N_\ai$, $R=R_\oi$. We set $t=N_\ai T$ and $r\to x$, and obtain
\begin{equation}\label{eq:MinkTorus}
    \ts{g} = - \dif t^2 + \dif x^2 + R_\oi^2\, \dif\ph^2\;.
\end{equation}
It describes the Minkowski metric with $2\pi$-periodic coordinate $\ph$, i.e., the compactified spacetime just discussed, but written in global inertial coordinates.

The metric \cref{eq:MinkTorus} can be regarded as a limit $\mBTZ\to1$ of the conical geometry \cref{eq:MinkConeBTZ}. Indeed, one has to perform the limit $c\to0$ with $\mBTZ=1-c^2$, $T=\frac{t}{c}$, and $R=R_\oi+c x$. It describes taking the cone with a vanishing vertex angle near the fixed circle $R=R_\oi$.

Summarizing, the generic vacuum static rotation-symmetric spacetime with a vanishing cosmological constant behaves in the asymptotic region as conical spacetime \cref{eq:MinkConeGeom}, which can also be written in the form \cref{eq:MinkConeBTZ}. The special case is a toroidal flat spacetime, which can be written either as \cref{eq:MinkRind} or \cref{eq:MinkTorus}.


\subsection{Anti-de~Sitter case, $\Lambda<0$}
\label{ssec:AsymptAdS}

We are mostly interested in solutions with a negative cosmological constant and AdS asymptotics. Therefore, we will be a bit meticulous in this section. 

The cosmological scale $\ell$ is in three dimensions introduced by ${\Lambda = -\frac{1}{\ell^2}}$ . The corresponding dark-energy source is given by 
\begin{equation}\label{eq:Lambdaepsp}
    \kap\eps = -\frac{1}{\ell^2} \qquad\qquad \kap p = \frac{1}{\ell^2}.
\end{equation}
Solving \cref{eq:fluidEq}, we get
\begin{equation}\label{eq:asympAdSexppm}
    \begin{aligned}
        R &= R_\oplus \exp\bigl({\tfrac{r}{\ell}}\bigr) + R_\ominus \exp\bigl({-\tfrac{r}{\ell}}\bigr)\;,\\
        N &= N_\oplus \exp\bigl({\tfrac{r}{\ell}}\bigr) + N_\ominus \exp\bigl({-\tfrac{r}{\ell}}\bigr)\;,
    \end{aligned}    
\end{equation}
and the constraint $\eps=p_r=p_\ph=\tfrac{1}{\ell^2}$ implies
\begin{equation}
    N_\oplus=k R_\oplus\;,\quad N_\ominus=k R_\ominus\;,\quad k=\const\;.
\end{equation}
We have $R\ge0$, $N\ge0$, and we expect the asymptotic region for $r\to\infty$. Therefore, $R_\oplus,N_\oplus>0$. 

This general solution splits into two qualitatively different generic cases. In the first case, $R$ vanishes for some $r=r_\oi$, and in the latter case, $R'$ has a zero for $r=r_\oi$. It is straightforward to check that these two cases differ by the sign of $R_\ominus$,
\begin{equation}
    \begin{aligned}
        R=0\quad  &\text{for}\; \exp\frac{2r_\oi}{\ell} = -\frac{R_\ominus}{R_\oplus} &\text{with}\quad R_\ominus<0\;,\\
        R'=0\quad &\text{for}\; \exp\frac{2r_\oi}{\ell} = +\frac{R_\ominus}{R_\oplus} &\text{with}\quad R_\ominus>0\;.
    \end{aligned}
\end{equation}
A simple redefinition $R_\oplus=R_\ai e^{-\frac{r_\oi}{\ell}}$, $R_\ominus=\mp R_\ai e^{\frac{r_\oi}{\ell}}$, and ${N_\ai=k R_\ai}$, followed by the shift $r\to r+r_\oi$ of the radial coordinate (effectively enforcing $r_\oi=0$) gives us
\begin{equation}\label{eq:asympAdSexp}
    \begin{aligned}
        R = R_\ai \frac12\Bigl[\exp{\Bigl(\frac{r}{\ell}\Bigr)} \mp \exp\Bigl({-\frac{r}{\ell}}\Bigr)\Bigr]
           \;,\\
        N = N_\ai \frac12\Bigl[\exp\Bigl({\frac{r}{\ell}}\Bigr) \pm \exp\Bigl({-\frac{r}{\ell}}\Bigr)\Bigr]
           \;.
    \end{aligned}
\end{equation}

The first case, when the spatial geometry has a semi-regular origin at $r=0$, yields \cref{eq:statrotmetricTrph} with
\begin{align}\label{eq:asympAdScon}
    R = c_\ai\ell \sinh{\tfrac{r}{\ell}}\;,\quad
    N = N_\ai \cosh{\tfrac{r}{\ell}}\;.
\end{align}
Here, we redefined the parameter $R_\ai\equiv c_\ai\ell$.
The second case corresponds to spatial wormhole geometry with the throat at $r=0$, i.e., the metric \cref{eq:statrotmetricTrph} with
\begin{align}\label{eq:asympAdSBH}
    R = R_\hor \cosh{\tfrac{r}{\ell}}\;,\quad
    N = N_\ai \sinh{\tfrac{r}{\ell}}\;.
\end{align}    
Here, we renamed $ R_\ai\equiv R_\hor$.

For completeness, we should include the case $R_\ominus=0$,
\begin{align}\label{eq:asympAdSspike}
    R = R_\ai \exp{\tfrac{r}{\ell}}\;,\quad
    N = N_\ai \exp{\tfrac{r}{\ell}}\;.
\end{align}
In this case, the spatial geometry has neither a semi-regular origin nor a wormhole throat. It represents an infinitely long spike with a vanishing vertex angle. Notice that the freedom in the choice of the origin of the radial coordinate $r$ has not been fixed in this case.

We will dedicate separate sections to the first two cases below. Although they have different physical interpretations, all cases can be written in the famous BTZ form\footnote{Our form of the BTZ metric differs from the original one by a shift of the mass parameter $\mBTZ$, namely ${1-\mBTZ = -M_\BTZ}$.} \cite{Banados:1992a},
\begin{equation}
    \label{eq:BTZ_metric}
	\ts{g}_\BTZ = -\left(1{-}\mBTZ{+}\tfrac{R^2}{\ell^2}\right)\dif T^2 
        + \frac{1}{1{-}\mBTZ{+}\tfrac{R^2}{\ell^2}}\, \dif R^2 
        + R^2\,\dif \ph^2 \;.
\end{equation}
The first case occurs for $\mBTZ<1$; the other occurs for $\mBTZ>1$. The special case corresponds to $\mBTZ=1$.

The metric \eqref{eq:statrotmetricTrph} with metric functions given by \eqref{eq:asympAdSexppm}, or split into separate cases 
\eqref{eq:asympAdScon}, \cref{eq:asympAdSBH}, and \cref{eq:asympAdSspike}, or, alternatively, the BTZ metric \eqref{eq:BTZ_metric}, can be considered as geometries with canonical AdS asymptotics. They represent a static, rotation-symmetric spacetime with only dark energy in the distant regions. Although there are three global types of these geometries, the circumference-radial coordinate $R$ grows exponentially with radial length $r$ in all cases. We will show that when ${R\propto\sinh\frac{r}{\ell}}$, the geometry extends to empty AdS with a semi-regular origin representing a point-like particle, and for ${R\propto\cosh\frac{r}{\ell}}$, it extends to a black hole solution.

\subsubsection*{AdS spacetime with a conical deficit}

Let us discuss geometry \eqref{eq:statrotmetricTrph} with the metric functions \eqref{eq:asympAdScon}. We invert the relation $R(r)$ and express the metric functions in terms of the coordinate $R$, 
\begin{subequations}\label{eq:rot_sym_R}
    \begin{gather}
        R'^2 = c_\ai^2 + \frac{R^2}{\ell^2}\;,\label{eq:RprofRPP}\\
        N^2 = N_\ai^2\left(1 + \frac{R^2}{\ell^2 c_\ai^2}\right)\;,\\
        N' = \frac{N_\ai}{c_\ai \ell^2} R\;.
    \end{gather}
\end{subequations}

For generic $c_\ai$, the geometry has a semi-regular origin at $r=0$ with a conical defect. Indeed, near the origin, the length of the orbit of the Killing vector $\ts{\partial}_\ph$ behaves as
\begin{equation}
    2\pi R\approx 2\pi c_\ai r\;,
\end{equation}
with $r$ being the proper radial distance from the origin. The constant $c_\ai$ describes a deviation from the regularity known as conicity. The regularity is restored for $c_\ai = 1$.

The solution with a general conicity can be constructed using the standard procedure that we have already seen in \Ref{ssec:AsymptMink}.
Let us consider the empty regular AdS given by $c_\ai=N_\ai=1$,
\begin{equation}
    \label{eq:AdSpolar}
	\ts{g}_\AdS = -\left(1{+}\tfrac{R^2}{\ell^2}\right)\dif T^2 
        + \frac{1}{1{+}\tfrac{R^2}{\ell^2}}\, \dif R^2 
        + R^2\,\dif \phi^2 \;.
\end{equation}
The spatial geometry at the slice $T=\const$,
\begin{equation}
    \ts{q} = \dif r^2 + \ell^2 \sinh^2\!\tfrac{r}{\ell}\, \dif \phi^2 \;,
\end{equation}
is a hyperbolic plane in polar coordinates.

We can now restrict the angular coordinate $\phi$ to an interval\footnote{ We will see that the physically most relevant case is $\phi_\oi\in(0,\pi)$, but the range can be meaningfully extended to $\phi_\oi\in(0,\infty)$.} $\phi\in(-\phi_\oi,\phi_\oi)$. Now, we identify surfaces ${\pm\phi_\oi}$, i.e., points with ${\phi=\pm\phi_\oi}$ and the same $T$ and $R$. In the hyperbolic surface $T=\const$, this is equivalent to cutting out a central angular wedge and gluing the newly formed edges. Similar to the flat plane, we obtain a hyperbolic cone; hence this defect is called conical.

In this restricted spacetime, we introduce new rescaled coordinates
\begin{equation}
    \label{eq:transform}
    \tilde{T} = \frac{1}{N_\ai}T \;,\qquad 
    \tilde{R} = \frac{\phi_\oi}{\pi}R \;,\qquad 
    \ph = \frac{\pi}{\phi_\oi}\phi \;.
\end{equation}
We thus restored the range of the angular coordinate to ${\ph\in(-\pi,\pi)}$ and, dropping the tildes, obtained the conical spacetime \eqref{eq:rot_sym_R} with generic $c_\ai=\frac{\phi_\oi}{\pi} $ and $N_\ai$. Of course, $N_\ai$ can be set to an arbitrary value by rescaling the Killing coordinate $T$. Typically, we set $N_\ai=1$, thus choosing $T$ to be a proper time at the origin.

Furthermore, one can replace the parameter $c_\ai$ (or, equivalently, $\phi_\oi$) with a new parameter $\mBTZ$ defined again by 
\begin{equation}
    1-\mBTZ \equiv  c_\ai^2 = \frac{\phi_\oi^2}{\pi^2}
\end{equation} 
to acquire the metric in the form \cref{eq:BTZ_metric}. The present case ${\mBTZ\in(-\infty,1)}$ can be interpreted as a point particle solution, identifying the particle of mass $\mBTZ$ with the history of the conical defect at the center. This solution has been known since the sixties (e.g. \cite{Staruszkiewicz:1963,Deser:1984a}). See \cite{Lukes:2026a} for its recent thorough description and extension. 

Alternatively, we can again introduce the mass parameter $\hat\locmass_\oi$, directly related to the angular deficit by \cref{eq:locmassconicity}. We will see below that this parameter can be obtained as a limit of the mass of the extended object simulating the point-like source.

We will now turn to the more interesting case $\mBTZ>1$.

\subsubsection{BTZ black hole}

Let us investigate the geometry \eqref{eq:statrotmetricTrph} with the metric functions \eqref{eq:asympAdSBH}. The circumference-radial coordinate $R$ extends from infinity down to a finite minimal value $R_\hor$ at $r=0$. Inverting the relation $R(r)$ and expressing the metric functions in terms of $R$ leads to
\begin{equation}\label{eq:BH_R}
    \begin{aligned}
        R'^2 &= \frac{R^2-R_\hor^2}{\ell^2}\;,\\
        N^2  &= N_\ai^2\frac{R^2-R_\hor^2}{R_\hor^2}\;,\\
        N' &= \frac{N_\ai}{R_\hor \ell} R
    \end{aligned}
\end{equation}

The solution \cref{eq:BH_R} describes spacetime only down to ${R=R_\hor}$, where both $R'$ and $N$ are zero. It indicates the presence of the horizon \cite{Banados:1992a}. The coordinates can be analytically continued into regions beyond the horizon and cover the whole spacetime\footnote{Different patches of coordinates $T,R,\ph$ can cover all parts of the spacetime except the horizon. One can also find globally smooth coordinates covering the entire spacetime, including the horizon.} \cite{Lukes:2026c,Banados:1992a,Banados:1992b}.

The spatial section ${T=\const}$ can be extended beyond the minimal value $R_\hor$, allowing $r\in\realn$. It glues two copies of spatial geometry at the orbit of minimal value $R_\hor$, representing the spatial geometry of the Einstein-Rosen bridge of the analytically extended black hole solution.

We can obtain this solution by a procedure analogous to the construction of the conical deficit in the previous section, see \cite{Lukes:2026c} for more details. One can start with the solution ${R_\hor=\ell}$, ${N_\ai=1}$ and with the relaxed restriction on the angular coordinate, allowing ${\phi\in\realn}$,
\begin{equation}
    \label{eq:AdSacc}
	\ts{g}_\AdS = -\left(\tfrac{R^2}{\ell^2}{-}1\right)\dif T^2 
        + \frac{1}{\tfrac{R^2}{\ell^2}-1}\, \dif R^2 
        + R^2\,\dif \phi^2 \;.
\end{equation}
Surprisingly, this solution also corresponds to the empty regular AdS spacetime. It is partially covered by the coordinates $T, r, \phi$ adapted to two Killing vectors. However, the static Killing vector $\ts{\partial}_T$ now has a bifurcation character, and the orbits of $\ts{\partial}_\phi$ are not compact. The spatial section ${T=\const}$ is a hyperbolic plane with the metric 
\begin{equation}
    \ts{q} = \dif r^2 + \ell^2 \cosh^2\!\tfrac{r}{\ell}\, \dif \phi^2 
\end{equation}
written in coordinates adjusted to the translation along the Killing vector $\ts{\partial}_\phi$ and the distance from the axis ${r=0}$. 

Now we cut the spacetime along two surfaces $\phi=\pm\phi_\oi$, such that ${2\pi R_\hor=2\phi_\oi\ell}$, and glue them together at the points with the same $T$ and $r$. If one were to perform this procedure on a flat plane, one would obtain a cylinder. However, in the hyperbolic plane, such a cylinder has a growing circumference in both directions $r\to\pm\infty$, thus describing the geometry of a spatial wormhole joining two asymptotic regions with hyperbolic geometry. When performing the gluing procedure in the full spacetime, one obtains the black hole spacetime with two asymptotic regions on both sides of the horizon \cite{Lukes:2026c}.

As in the previous section, we rescale the coordinates as \eqref{eq:transform}, set ${N_\ai=1}$, and introduce the mass parameter
\begin{equation}
    \mBTZ-1 \equiv  \frac{R_\hor^2}{\ell^2} = \frac{\phi_\oi^2}{\pi^2}\;,
\end{equation}
obtaining the BTZ metric \eqref{eq:BTZ_metric}, now with $\mBTZ>1$. 

The BTZ black hole solution can be extended with rotation, acceleration, and charge \cite{Banados:1992a,Banados:1992b}.


\subsection{De~Sitter case, $\Lambda>0$}
\label{ssec:AsymptdS}

For ${\Lambda =\frac{1}{\ell^2}>0}$, we find, by arguments similar to the previous case, the geometry \cref{eq:statrotmetricTrph} with
\begin{equation}\label{eq:asympdScon}
    N=N_\oi \cos\tfrac{r}{\ell}\;,\quad R=c_\oi \ell \sin\tfrac{r}{\ell}\;,
\end{equation}
where we fixed the semi-regular origin at $r=0$ and $r=\pi\ell$. Both origins have conical defects with the same conicity $c_\oi$. $N_\oi$ represents the lapse at the origins and can be set arbitrarily; typically $N_\oi=1$. The lapse vanishes at $r=\frac\pi2\ell$, which is the Killing horizon of $\ts{\partial}_T$. 

The metric represents the de~Sitter universe in static coordinates with a conical deficit in two antipodal worldlines. Physically, it describes two point-like particles in the de~Sitter background. The spatial section $T=\const$ has geometry
\begin{equation}
    \ts{q} = \dif r^2 +c_\oi^2\ell^2\sin^2\!\tfrac{r}{\ell}\dif\ph^2
\end{equation}
of the homogeneous sphere with conical defects at two antipodal poles. 

In terms of the circumference coordinate $R$, setting $N_\oi=c_\oi$ and $1-\mBTZ=c_\oi^2$, the metric takes the BTZ-like form
\begin{equation}
    \label{eq:dScon}
	\ts{g} = -\left(1{-}\mBTZ{-}\tfrac{R^2}{\ell^2}\right)\dif T^2 
        + \frac{1}{1{-}\mBTZ{-}\tfrac{R^2}{\ell^2}}\, \dif R^2 
        + R^2\,\dif \ph^2 \;,
\end{equation}
with $T\in\realn$, $R\in(0,\sqrt{1-\mBTZ}\ell)$, $\ph\in(-\pi,\pi)$, cf.~\eqref{eq:BTZ_metric}.

Since the spatial section is compact, we do not have spatial infinity. Nevertheless,  the spacetime geometry approximates the temporal asymptotic of a generic solution with matter content approaching the cosmological term $\Lambda>0$ in the far past and future. However, these regions are not covered by the static regions studied here.
    \section{Simple solutions}
\label{sec:matter}

Now, we introduce two simple matter models: the incoherent dust and the incompressible fluid \cite{Einstein:1922}. After that, we discuss the static rotation-symmetric thin shell~\cite{Israel:1966}.


\subsection{Incoherent dust}
\label{ssec:dust}

The incoherent dust is characterized by an energy density $\eps_\dst$ and vanishing pressure. Including the cosmological constant, we have the stress-energy tensor \cref{eq:statrotT}
\begin{equation}
    \kap\eps = \Lambda + \kap\eps_\dst\;,\quad
    \kap p = - \Lambda\;.
\end{equation}
Temporal and radial angular components \cref{eq:fluidEq} of the Einstein equations thus give
\begin{subequations}
\begin{gather}
    R'' = -(\Lambda +\kap\eps_\dst)R\;,\label{eq:dustReq}\\ 
    N'' = - \Lambda N\;.\label{eq:dustNeq}
\end{gather}
\end{subequations}
The radial component \cref{eq:fluidisotropy} implies
\begin{equation}\label{eq:dustIsotropy}
    N' = k R\;,
\end{equation}
with $k$ constant and finite.\footnote{We do not admit $k=\infty$ since we assume that $R$ can achieve nonvanishing values.} 

Since ${\Lambda=\const}$ and assuming ${k\neq0}$, the derivative of \cref{eq:dustNeq} compared to \eqref{eq:dustReq} implies ${\eps_\dst=0}$, i.e., no dust. The only other option is ${k=0}$, which means that ${N=N_\oi}$ is constant. But it also requires the vanishing cosmological constant, ${\Lambda=0}$. 

The energy density $\eps_\dst$ can then be arbitrary and is related to an arbitrary choice of the metric function $R$ as
\begin{equation}\label{eq:epsdustR}
    \kap \eps_\dst = -\frac{R''}{R}\;.
\end{equation}
It demonstrates the absence of gravitational attraction between non-interacting matter in 2+1 dimensions. Each part of the dust is moving along a geodesic: along the orbit of the static Killing vector. Indeed, the magnitude of the static Killing vector given by the lapse is constant, and since the acceleration of the orbits is proportional to the gradient of the lapse, it vanishes. Particles of the incoherent dust are thus moving independently along the static Killing vector. However, their density distribution deforms the spatial geometry. The Gauss curvature of the spatial geometry is given exactly by the dust energy density, cf.~\cref{eq:EinSccur}. This means that one can prescribe an arbitrary spatial rotation-symmetric geometry by giving the dependence of the circumference radius $R(r)$ on the radial coordinate, and \cref{eq:epsdustR} provides the necessary dust density for a corresponding static spacetime geometry. 

Let us discuss a special case of homogeneous dust, i.e., constant positive energy density ${\eps_\dst=\eps_\oi>0}$. Requiring also a regular origin at ${r=0}$, the solution for the metric functions $R$ and $N$ is
\begin{equation}\label{eq:dust_r}
    \begin{aligned}
        R  = R_\oi \sin{\frac{r}{R_\oi}}\;,\quad
        R' = \cos{\frac{r}{R_\oi}}\;,\\
        N  = N_\oi \;,\quad
        N' = 0\;. \qquad
    \end{aligned}
\end{equation}
The scale $R_\oi$ is given by $R_\oi^2 = \frac{1}{\kap \eps_\oi}$. After promoting $R$ to a coordinate, we get $R\in(0,R_\oi)$,
\begin{equation}
    \label{eq:dust_R}
    R' = \sqrt{1-\kap\eps_\oi R^2}\;,
\end{equation} 
Setting ${N_\oi=1}$, and ${\tht=\frac{r}{R_\oi}}$, the metric \eqref{eq:statrotmetricTrph} takes the form
\begin{equation}\label{eq:EinstStUniv}
  \begin{split}
  \ts{g} &= - \dif T^2 + R_\oi^2 \bigl(\dif \tht^2 + \sin^2\tht\, \dif\ph^2\bigr) \\
         &= -\dif T^2 +\frac{\dif R^2}{1-\kap\eps_\oi R^2} + R^2\dif\ph^2\;.
  \end{split}
\end{equation}
For ${\tht\in(0,\pi)}$, this is the static homogeneous sphere of radius $R_\oi$ with a global proper time $T$, i.e., the three-dimensional version of the Einstein static universe. It has two ``origins'' -- poles at ${\tht=0}$ and ${\tht=\pi}$. Notice that in three dimensions, the Einstein static universe has a different matter content: it is filled with homogeneous dust without pressure instead of the cosmological term in four dimensions.

Note that while the use of trigonometric functions in \cref{eq:dust_r} relied on the assumption ${\eps_\oi > 0}$, the relation \cref{eq:dust_R} and the second form of the metric hold for all values of energy density. For negative density, the trigonometric functions must be replaced by hypergeometric ones. We obtain a static hyperbolic geometry with universal proper time $T$, a hyperbolic version of the Einstein static universe. 

To summarize, the static dust solution is possible only in the absence of the cosmological constant. The energy density $\eps_\dst$ can then be arbitrary. It is worth noting that for the static rotationally symmetric ansatz, the energy density $\kap\eps$ is equal to the Gauss curvature $\frac12\spsc$ of the geometry on the time slice $T=\const$. For homogeneous dust, we thus obtain a spatially homogeneous sphere.


\subsection{Incompressible perfect fluid}
\label{ssec:fluid}

Next, let us assume the ansatz with a constant positive density ${\eps=\eps_\oi>0}$ and variable pressure $p(r)$. The constant energy density is a naive form of the fluid's equation of state. However, it allows us to find a simple model of a compact star-like object in empty spacetime.\footnote{\label{fnt:fluidLambda}It allows the presence of the cosmological term, in which case we have ${\kap\eps=\Lambda+\kap\eps_\fld=\const}$ and ${\kap p(r) = -\Lambda + \kap p_\fld(r)}$, cf.~\cref{eq:modEpsP} in \ref{ssec:FluidAdS}.} We again require a regular origin at ${r=0}$, which, together with \cref{eq:fluidEq,eq:fluidisotropy}, implies
\begin{equation}\label{eq:fluid_r}
    \begin{gathered}
        R  = \frac{1}{\sqrt{\kap\eps_\oi}} \sin\bigl(\sqrt{\kap\eps_\oi}r\bigr)\;,\\
        R' = \cos\bigl(\sqrt{\kap\eps_\oi}r\bigr)\;,\\
        N  = N_\oi\left( 1 + \frac{\kap p_\oi}{\kap \eps_\oi} \Bigl(1-\cos\bigl(\sqrt{\kap\eps_\oi}r\bigr)\Bigr) \right)\;,\\
        N' = N_\oi \frac{\kap p_\oi}{\sqrt{\kap \eps_\oi}}\sin\bigl(\sqrt{\kap\eps_\oi}r\bigr)\;,
    \end{gathered}
\end{equation}
with the pressure given by 
\begin{equation}
    \kap p = \frac{\kap p_\oi \cos(\sqrt{\kap\eps_\oi}r)}
         {1+\frac{\kap p_\oi}{\kap \eps_\oi}\bigl(1-\cos(\sqrt{\kap\eps_\oi}r)\bigr)}\;.
\end{equation}
Here, $N_\oi$ and $p_\oi$ are the values of the lapse and the pressure at the origin. 

Inverting the relation $R(r)$ and eliminating $r$, we acquire
\begin{subequations}
    \label{eq:fluid_R}
    \begin{gather}
        R' = \sqrt{1 - \kap\eps_\oi R^2} \;,\label{eq:fluid_RderR}\\
        N  = N_\oi\Bigl(1+\frac{\kap p_\oi}{\kap \eps_\oi}\bigl(1 - R' \bigr)\Bigr)\;,\label{eq:fluid_Rlapse}\\
        N' = N_\oi\kap p_\oi R\;,\label{eq:fluid_Rlapseder}
    \end{gather}
\end{subequations}
and
\begin{equation}\label{eq:fluidpressureR}
   \kap p  = \frac{\kap p_\oi R'}{1+\frac{\kap p_\oi}{\kap \eps_\oi}(1-R')} \;.
\end{equation}

Near the origin, the spatial geometry is again a homogeneous sphere with the latitudinal angular coordinate ${\tht=\sqrt{\kap\eps_\oi} r}$. However, the static time differs from the proper time of static observers by the lapse function, which now depends on the radial coordinate. The pressure is a decreasing function with a maximum $p_\oi$ at the origin, and it vanishes at $\tht=\frac\pi2$. To avoid negative pressures, one has to cut the solution at  $\tht=\tht_*\le\frac\pi2$ and join it with a solution for another matter content. We will discuss this in \ref{ssec:FluidAdS}.

If we allowed negative energy density $\eps_\oi<0$, we would need to use hyperbolic instead of trigonometric functions.


\subsection{Static thin shell}
\label{ssec:shell}

In \cref{sec:combine}, we will combine the inner matter solutions with suitable outer empty asymptotic solutions. The surface where the solutions are joined can, in principle, contain additional singular matter content, depending on the regularity of the metric function. A general form of the junction conditions was discussed by W. Israel in \cite{Israel:1966}.

Two parts of the spacetime  $\dom{M}_+$ and $\dom{M}_-$ can be glued together along the common surface $\Sigma$ if the restriction of the metrics $\ts{g}_\pm$ on $\Sigma$ is the same, 
\begin{equation}
    \ts{h}\equiv\ts{g}_-|_\Sigma = \ts{g}_+|_\Sigma \;,
\end{equation}
where we named the induced metric $\ts{h}$.

Let us denote the inner normals of both domains on the boundary $\Sigma$ as $\ts{n}_\pm$ and a common normal ${\ts{n}\equiv - \ts{n}_- = \ts{n}_+}$. The extrinsic curvature $\ts{K}_\pm$ of the geometry on both sides is defined using this common normal,
\begin{equation}
    \ts{K}_\pm = \ts{h}\cdot\ts{\nabla}_{\!\pm} \ts{n} \;,
\end{equation}
with Levi-Civita derivatives $\ts{\nabla}_{\!-}$ and $\ts{\nabla}_{\!+}$ defined in $\dom{M}_-$ and  $\dom{M}_+$, respectively.

The jump in the extrinsic curvature on the hypersurface $\Sigma$, ${[\ts{K}] \equiv (\ts{K}_+ - \ts{K}_-)|_\Sigma}$, characterizes a distributional curvature\footnote{Note that the direction of the normal $\ts{n}$ determines the sign of the extrinsic curvatures and also the direction from $\dom{M}_-$ to $\dom{M}_+$. So not only would the change $\ts{n}\to-\ts{n}$ switch the signs of the extrinsic curvatures, but it would also switch the roles of $\ts{K}_+$ and $\ts{K}_-$ in $[\ts{K}]$. Therefore, the quantity $[\ts{K}]$ is invariant under a change of the orientation of the normal.} localized on $\Sigma$. Plugging into the Einstein equations, W.~Israel \cite{Israel:1966} showed that the hypersurface supports a thin massive shell with a distributional stress-energy tensor 
\begin{equation}\label{eq:Tshell}
   \Terg=\ts{S}\delta_\Sigma\;,
\end{equation} 
where
\begin{equation}\label{eq:Sshell}
   \kap \ts{S} =  - [\ts{K}] + \ts{h}\,\Tr([\ts{K}])\;.
\end{equation}

Let us determine the stress-energy tensor arising from the identification of two static-rotation-symmetric domains along a static-rotation-symmetric boundary $\Sigma$ at ${r=r_*}$. Starting with metrics \cref{eq:statrotmetricTRph} on both sides, the induced metric $\ts{h}$ and the common normal $\ts{n}$ orthogonal to $\Sigma$ are
\begin{equation}
    \begin{gathered}
        \ts{h} = -N^2 \dif T^2 + R^2 \dif \ph^2\;,\\
        \ts{n} = \ts{\partial}_{r}\;.
    \end{gathered}
\end{equation}
Strictly speaking, we should distinguish the coordinates $T_\pm$ and $\ph_\pm$ in both domains, as well as the metric functions $N_\pm$ and $R_\pm$. However, the Killing tensor $\ts{\partial}_{\ph_\pm}$ and the corresponding coordinates $\ph_\pm$ are uniquely normalized by the periodicity condition, and therefore, they must be continuous on $\Sigma$. Similarly, we can normalize $T_\pm$ in such a way that it matches on $\Sigma$. We can thus use globally defined coordinates $T$ and $\ph$, which can be regarded as smooth on $\Sigma$. The continuity of the metric then also gives the continuity of $N$ and $R$ on $\Sigma$. However, the derivatives $N'$ and $R'$ do not have to be continuous on $\Sigma$, and we naturally define jumps $[N']=(N'_+-N'_-)|_\Sigma$ and $[R']=(R'_+-R'_-)|_\Sigma$.

The extrinsic curvature of the hypersurface ${r=\const}$ is
\begin{equation}
    \ts{K} = \ts{h}\cdot\nabla\ts{n} = -NN'\dif T^2 + RR'\dif\ph^2\;
\end{equation}
and the jump at $\Sigma$ reads
\begin{equation}
    [\ts{K}] = -N[N']\dif T^2 + R[R']\dif\ph^2\;.
\end{equation}
We thus obtain the surface stress-energy tensor \cref{eq:Sshell}
\begin{equation}
    \kap\ts{S} = -\frac{[R']}{R}N^2\dif T^2 + \frac{[N']}{N}R^2\dif \ph^2\;.
\end{equation}
It can be interpreted as a thin shell\footnote{In 2+1 dimensions, the thin shell is localized on a two-dimensional time-like hypersurface $\Sigma$, which represents the history of a one-dimensional spatial object. It plays both the role of a domain wall and a cosmic string. Therefore, we speak about linear energy density $\eTS$ and tension $\pTS$.} localized on the hypersurface $\Sigma$ with a linear energy density $\eTS$ and tangential tension $\pTS$ satisfying
\begin{equation}\label{eq:shellepsp}
    \kap\eTS = -\frac{[R']}{R}\;,\quad \kap \pTS = \frac{[N']}{N}\;.
\end{equation}
The dimensionless variants of these read\footnote{Here, we need a cosmological scale $\ell$ for introducing the dimensionless quantities.}
\begin{equation}
     \ehTS=\kap\eTS\ell\;,\quad \phTS=\kap\pTS\ell\;.
\end{equation}

Notice that the thin shell is not a subcase of the isotropic fluid: the radial pressure is irrelevant and not equal to the tangential one. One cannot thus use the analogy of condition \cref{eq:fluidisotropy}.

A particular physical model of a thin shell should be specified by an equation of state that relates the linear density to the tension. 

The incoherent dust localized on the shell corresponds to vanishing tension, $\pTS=0$. 

A natural equation of state is the condition that the stress-energy tensor is proportional to the hypersurface metric
\begin{equation}
    \ts{S} = -\eTS\,\ts{h}\;,
\end{equation}
It can be related to the Nambu-Goto action for the shell \cite{Barrabes:1991,Vilenkin:2000}, and therefore, we call it the Nambu-Goto equation of state. In terms of density and tension, it means
\begin{equation}\label{eq:shellCS}
    \eTS = -\pTS\;.
\end{equation}

Another reasonable equation of state is that of radiation, i.e., the shell with the world-sheet stress-energy tensor of vanishing trace,
\begin{equation}
    \Tr\ts{S} = 0\;,
\end{equation}
For a two-dimensional shell in 2+1 spacetime dimensions, it means 
\begin{equation}\label{eq:shellrad}
    \eTS = \pTS\;.
\end{equation}
Such a matter can be realized as two counter-rotating congruences of photons tangent to the shell.

Clearly, the requirement of gluing both domains without additional matter at their boundary would imply the continuity of $N'$ and $R'$ on $\Sigma$.
    \section{Integral mass}
\label{sec:mass}

To quantify the mass of a static-rotation symmetric spacetime, we introduce two simple definitions. Both are based on the existence of a static frame and integrate over the static spatial slice.


\subsection{Local mass}

First, we introduce a plain integral of the energy density~$\eps$ over a chosen domain. In a rotation-symmetric spacetime, we consider the domain $\dom{D}$ defined by the condition $r\in({r_\iix},{r_\oix})$, representing an annulus. We define the \emph{local mass}
\begin{equation}
    \label{eq:locmass}
    \locmass_{\iix}^{\oix} = \int_{\dom{D}} \ts{u}\cdot\Ein\cdot \ts{dS} = \int_{\dom{D}} \eps\, dS\;,
\end{equation}
where $\ts{u}$ is a timelike unit vector (the velocity of static observers) and $\ts{dS} = \ts{u}\,dS$ with $dS=R\,drd\ph$ being an area element defined by the spatial metric ${\ts{q}=\dif r^2+R^2\dif \ph^2}$. Employing \eqref{eq:fluidEq} we obtain
\begin{equation}
    \label{eq:locmassdomD}
    \begin{split}
        \kap\locmass_{r_\iix}^{r_\oix} &= -\int_{\dom{D}} \frac{R''}{R}\, R\, dr d\ph
        = -2\pi\int_{r_\iix}^{r_\oix} R'' dr \\
        &= 2\pi\bigl(R'|_{r_\iix}-R'|_{r_\oix}\bigr)\;,
    \end{split}
\end{equation}

At first glance, it can be surprising that the result does not depend on the details of the metric functions inside the annulus, it depends just on the value $R'$ at its boundary. This can be elucidated by the observation that the time-time component $\Ein_{\perp\perp}$ of the Einstein tensor (with respect to a normalized normal) is equivalent to the scalar curvature $\spsc$ of the spatial metric, cf.~\cref{eq:EinSccur}. From the Gauss-Bonnet theorem, it follows that (since there are no point-wise changes in the tangent vector along the boundaries of the annulus and its Euler characteristic is~0)
\begin{equation}
    \label{eq:loc_mass_result}
    \kap\locmass_{r_\iix}^{r_\oix}=\Delta\Phi_\oix - \Delta\Phi_\iix\;,
\end{equation}
where $\Delta\Phi$ is the angle deficit of the parallel transported vector along the inner or outer boundary of our annulus. The statement is true not only for a disk but also for any domain bounded by two simple loops. The local mass in the domain can thus be measured by parallel transport around its boundaries. It is an explicit demonstration of the relationship between curvature and parallel transport. 

We can observe that the formula \cref{eq:locmassdomD} also holds for matter, including thin shells. One can check it directly by splitting the spacetime into the domain $\dom{D}_-$ below the shell, the shell, and the domain $\dom{D}_+$ above the shell. The contribution of the thin shell follows from the distributional form of the energy density when restricted to the time slice $T=\const$, cf.\ \cref{eq:Tshell}, \cref{eq:Sshell}, and \cref{eq:shellepsp},
\begin{equation}\label{eq:shelleps}
    \kap\eps_{\mathrm{shell}} = \kap \eTS\, \delta(r|r_*) = -\frac{[R']}{R}\, \delta(r|r_*)\;,
\end{equation}
\begin{equation}
    \kap\locmass_{\mathrm{shell}} = \kap \int_{r_-}^{r_+}\eps_{\mathrm{shell}}\, dS 
      = -2\pi [R']
      \;.
\end{equation}
The local mass thus reads
\begin{equation}\label{eq:locmasshellsplit}
\begin{split}
    &\kap\locmass_{\iix}^{\oix}=\kap\locmass_{\dom{D}_-}+\kap\locmass_{\mathrm{shell}}+\kap\locmass_{\dom{D}_+}\\
      &\quad= 2\pi\bigl(R'|_{r_\iix}{-}R'|_{r_-}+R'|_{r_-}{-}R'|_{r_+}+R'|_{r_+}{-}R'|_{r_\oix}\bigr)\\
      &\quad= 2\pi\bigl(R'|_{r_\iix}-R'|_{r_\oix}\bigr)\;.
\end{split}\raisetag{3ex}
\end{equation}

For a regular origin, one can shrink the inner boundary to an infinitesimal loop around the origin, $r_\iix\to r_\oi$, with $\Delta\Phi_\iix = -2\pi$ and $R'|_{r_\oi} = 1$; therefore,
\begin{equation}\label{eq:locmassuptorout}
    \kap\locmass^{r_\oix}
    = 2\pi + \Delta\Phi_\oix = 2\pi(1-R'|_{r_\oix})\;.
\end{equation}

Finally, just for the simplification of various relations, we introduce a rescaled mass parameter\footnote{Here, we do not need a cosmological scale for obtaining the dimensionless quantities. But we use a customary numerical factor $2\pi$ related to the circumference (area) of the 1-dimensional circle (sphere) in two spatial dimensions.}
\begin{equation}\label{eq:locmasshat}
    \hat\locmass = \frac{\kap\locmass}{2\pi}\;.
\end{equation}
For example, the mass formula \cref{eq:locmassdomD} then reads
\begin{equation}
    \label{eq:locmassdomDhat}
    \hat\locmass_{r_\iix}^{r_\oix} = R'|_{r_\iix}-R'|_{r_\oix} \;.
\end{equation}


\subsection{Killing mass}

Another useful characterization of the matter content is the \emph{Killing mass} $\MKV$ associated with the Killing vector $\ts{\partial}_T$. The Killing mass inside a rotationally symmetric domain $\dom D$ as above is 
\begin{equation}
	\label{eq:KVmassdomD}
	\kap \MKV_{r_\iix}^{r_\oix} = \int_{\dom{D}} \ts{\partial}_T\cdot\Ein\cdot \ts{dS}\;,
\end{equation}
Employing again \eqref{eq:statrotEin} and \eqref{eq:fluidEq}, we obtain
\begin{equation}\label{eq:KVmassdomD1}
    \begin{split}
        \kap\MKV_{r_\iix}^{r_\oix} &= -\int_{\dom{D}} \frac{R''}{R}\,N\, R\, dr d\ph
                                   = -2\pi\!\int_{r_\iix}^{r_\oix} \!\!\!R'' N\,dr\\
                                   &= -2\pi\int_{r_\iix}^{r_\oix} \biggl( (R' N)'-\frac{N'}{R}\,\frac12(R^2)'\biggr)dr \;.
    \end{split}\raisetag{7.5ex}
\end{equation}
Using the isotropy condition \eqref{eq:fluidisotropy}, we can take $\frac{N'}{R}$ out of the integral of the second term, yielding 
\begin{equation}
    \label{eq:KVmassdomD2}
    \begin{split}
            \kap\MKV_{r_\iix}^{r_\oix}
            &= 2\pi\Bigl[\frac12\, N' R - R' N\Bigr]_{r_\iix}^{r_\oix} \;.
    \end{split}
\end{equation}

Since we have used the isotropy condition \eqref{eq:fluidisotropy}, the final formula \cref{eq:KVmassdomD2} is not guaranteed for matter that includes a thin shell. In such a case, we have to return to \cref{eq:KVmassdomD1}, and with the help of \cref{eq:shelleps} we obtain
\begin{equation}\label{eq:MKVshell}
    \kap\MKV_{\mathrm{shell}} = \kap \int_{r_-}^{r_+}\eps_{\mathrm{shell}} N \, dS
      = -2\pi [R'] N_* \;.
\end{equation}
Splitting the spacetime as in Eq.~\cref{eq:locmasshellsplit}, we obtain a modification of formula \cref{eq:KVmassdomD2} as
\begin{equation}\label{eq:MKVshellsplit}
\begin{split}
    &\kap\MKV_{\iix}^{\oix}=\kap\MKV_{\dom{D}_-}+\kap\MKV_{\mathrm{shell}}+\kap\MKV_{\dom{D}_+}\\
      &\quad= 2\pi\Bigl(\bigl[\tfrac12\, N' R - R' N\bigr]_{r_\iix}^{r_-}
        - \bigl[R' N\bigr]_{r_-}^{r_+}\\
        &\mspace{180mu}+ \bigl[\tfrac12\, N' R - R' N\bigr]_{r_+}^{r_\oix}\Bigr)\\
      &\quad= 2\pi\bigl[\tfrac12\, N' R - R' N\bigr]_{r_\iix}^{r_\oix}
        - \pi \bigl[N'\bigr] R_*\;.
\end{split}\raisetag{3ex}
\end{equation}
We see that it gets a contribution from a non-regularity of $N'$. But since we used the equations of motion and the isotropy of the fluid surrounding the shell, it is sensitive to the non-regularity of $[R']$ too, cf.~\cref{eq:MKVshell}.

For a disk with a regular origin, we can again take the limit of the inner boundary to the origin and obtain
\begin{equation}
    \label{eq:KVmassdomDregorg}
    \kap \MKV^{r_\oix}
    = {2\pi}\Bigl(\frac12 N' R - R' N + N_\oi\Bigr)\Big\vert_{r_\oix} \;.
\end{equation}
The result is expressed just in terms of values of the metric functions and their derivatives\footnote{Recall that the prime is the derivative with respect to $r$. If one gets the metric in the form \eqref{eq:statrotmetricTRph}, functions $R'(R)$ and $N(R)$ are directly available. For $N'(R)$ one can use $N'=\frac{dN}{dR}\,R'$.} on the boundary.

Again, we introduce a rescaled version of the Killing mass,
\begin{equation}\label{eq:MKVhat}
    \hat\MKV = \frac{\kap\MKV}{2\pi}\;.
\end{equation}


\subsection{Mass for asymptotic geometries}

For conical Minkowski geometry \cref{eq:MinkConeGeom} we have $R'=c_\ai$, $N'=0$, and the mass formulae \cref{eq:locmassdomD} and \cref{eq:KVmassdomD2} yield
\begin{gather}
    \hat\locmass_{r_\iix}^{r_\oix} = 0 \;,\\
    \hat\MKV_{r_\iix}^{r_\oix} = 0 \;.
\end{gather}
Of course, the spacetime does not contain any matter inside the annulus between $r_\iix$ and $r_\oix$. Since the spacetime does not have a regular origin, we should not apply the formulae \cref{eq:locmassuptorout} or \eqref{eq:KVmassdomDregorg}. However, in the following, we want to argue that if we regularize the matter distribution near the origin and then squeeze such a distribution to an infinitesimal region near the origin, the formulae \cref{eq:locmassuptorout} or \eqref{eq:KVmassdomDregorg} correctly describe the total mass in the spherical region, namely
\begin{gather}
    \hat\locmass^{r_\oix} = 1-c_\ai =\frac{\Delta\Phi}{2\pi}\;,\label{eq:MinkPPlocmass}\\
    \hat\MKV^{r_\oix} = N_\oi (1-c_\ai) = N_\oi \hat\locmass^{r_\oix}\;.\label{eq:MinkPPMKV}
\end{gather}
Here, ${\Delta\Phi = 2\pi-2\phi_\oi}$ is the asymptotic angular deficit, cf.\ \cref{eq:locmassconicity}.

In the conical AdS case \cref{eq:rot_sym_R}, the local mass \cref{eq:locmassuptorout} reads
\begin{equation}\label{eq:locmassconAdSinout}
\begin{split}
    \hat\locmass_{r_\iix}^{r_\oix} 
    &= c_\ai \bigl( \cosh\frac{r_\iix}{\ell} - \cosh\frac{r_\oix}{\ell}\bigr)\\
    &= \sqrt{c_\ai^2+\frac{R_\iix^2}{\ell^2}} -\sqrt{c_\ai^2+\frac{R_\oix^2}{\ell^2}} \;,
\end{split}
\end{equation}
We observe that it decreases as $r_\oix$ increases. Of course, it includes the contribution of dark energy, which is negative in the asymptotic region. Naive application of \cref{eq:locmassuptorout} gives 
\begin{equation}
    \hat\locmass^{r_\oix} 
    = 1- c_\ai \cosh\frac{r_\oix}{\ell}
    = 1 -\sqrt{c_\ai^2+\frac{R_\oix^2}{\ell^2}} \;.
\end{equation}
Comparing with \cref{eq:locmassconAdSinout} for ${r_\iix=0}$, namely
\begin{equation}
    \hat\locmass_{r_\iix}^{r_\oix}  = \hat\locmass_\oi + \hat\locmass_{r=0}^{r_\oix}\;.
\end{equation}
It suggests the presence of a point particle at the origin with mass 
\begin{equation}\label{eq:locmasscontip}
   \hat\locmass_\oi=1-c_\ai\;. 
\end{equation}
It corresponds to the conical contribution \cref{eq:MinkPPlocmass} in the flat case. However, we should stress that relation \cref{eq:locmassuptorout} assumes a regular origin and, strictly speaking, is not applicable in this case. We will justify this result in the discussion of models of matter with regular origin in \cref{sec:combine}.

Analogously, for the Killing mass \cref{eq:KVmassdomD2} we get
\begin{equation}\label{eq:MKVconAdSinout}
\begin{split}
    \hat\MKV_{r_\iix}^{r_\oix} 
    &= \frac12 N_\ai c_\ai \Bigl( \sinh^2\frac{r_\iix}{\ell} - \sinh^2\frac{r_\oix}{\ell}\Bigr)\\
    &= \frac12\frac{N_\ai}{c_\ai}\Bigl(\frac{R_\iix^2}{\ell^2} -\frac{R_\oix^2}{\ell^2}\Bigr) \;,
\end{split}
\end{equation}
and
\begin{equation}
\begin{split}
    \hat\MKV^{r_\oix} 
    &= N_\ai\Bigl(1- c_\ai -\frac12 c_\ai\sinh^2\frac{r_\oix}{\ell}\Bigr)\\
    &= N_\ai(1-c_\ai) + \hat\MKV_{r=0}^{r_\oix} \;,
\end{split}
\end{equation}
suggesting the point particle of the Killing mass ${\hat\MKV_\oi = N_\ai(1-c_\ai)}$.

Finally, for AdS black hole asymptotics \cref{eq:BH_R}, the local mass \cref{eq:locmassuptorout} gives
\begin{equation}\label{eq:locmassBHAdSinout}
\begin{split}
    \hat\locmass_{r_\iix}^{r_\oix} 
    &= \frac{R_\hor}{\ell} \bigl( \sinh\frac{r_\iix}{\ell} - \sinh\frac{r_\oix}{\ell}\bigr)\\
    &= \frac{1}{\ell}\bigl(\sqrt{R_\iix^2-R_\hor^2} -\sqrt{R_\oix^2-R_\hor^2}\bigr) \;.
\end{split}
\end{equation}
It decreases again as $r_\oix$ increases, owing to dark energy. Its contribution above the horizon is 
\begin{equation}\label{eq:locmassBHAdShorout}
    \hat\locmass_{r_\hor}^{r_\oix} 
    = -\frac{R_\hor}{\ell} \sinh\frac{r_\oix}{\ell}\\
    = -\frac{1}{\ell}\sqrt{R_\oix^2-R_\hor^2} \;.
\end{equation}
The Killing mass is given by 
\begin{equation}\label{eq:MKVBHAdSinout}
\begin{split}
    \hat\MKV_{r_\iix}^{r_\oix} 
    &= \frac12 N_\ai \frac{R_\hor}{\ell} \Bigl( \cosh^2\frac{r_\iix}{\ell} - \cosh^2\frac{r_\oix}{\ell}\Bigr)\\
    &= \frac12\frac{N_\ai\ell}{R_\hor}\Bigl(\frac{R_\iix^2}{\ell^2} -\frac{R_\oix^2}{\ell^2}\Bigr) \;.
\end{split}
\end{equation}
    \section{Models of a compact object}
\label{sec:combine}

Our goal is to investigate various models for a point-like source. We approximate the singular point-like sources with matter regular at the origin. Namely, we consider a static, spherically symmetric matter distribution confined to a ball of circumference radius $R_*$, surrounded by $\Lambda$-vacuum asymptotic geometry. We focus mostly on the object in AdS asymptotics. A dust, fluid, and dark-energy bubble will constitute the matter content within the central ball. In general, the matter must be shielded by a thin spherical shell to compensate for a difference in radial pressure. We consider the Nambu-Goto and radiation shells. We find the characterization of the asymptotic geometry in terms of the parameters of the matter model. 

To study the point-like source limit, we also need the inverse relation: to determine the matter model parameters for a given asymptotic geometry. Next, we perform the limit of the negligible size $R_*$ of the central object with the asymptotic geometry fixed. The limit should not be considered a physical dynamic process but merely an improving approximation of a point-like singular source. When considering the limit ${R_*\to0}$ and AdS asymptotics, we have to restrict to the conical case \cref{eq:asympAdScon}. The black hole asymptotics have a minimal circumference radius $R_\hor$, and the limit ${R_*\to0}$ does not make sense when $R_\hor>0$ is fixed.

We are mostly interested in the limits of the local and Killing masses of the central object. However, the masses associated with entire spacetimes with AdS asymptotics are negatively infinite because they include the dark-energy contribution. We need to disregard this dark-energy contribution outside the matter object. Fortunately, since the geometry is locally fully determined by the matter distribution, we know that the geometry outside the central object must be the $\Lambda$-vacuum asymptotic geometry, and we know the contribution of dark energy to the local and Killing mass in this geometry. Removing the dark-energy contribution outside the matter object amounts to calculating the mass just above the object's radius. We thus use formulas \cref{eq:locmassuptorout} and \cref{eq:KVmassdomDregorg} up to the radial coordinate $r_+$ just above the matter ball, including the shell on its surface. After that, we perform the limit of negligible size of the object, keeping the asymptotic geometry fixed.  

We will demonstrate that the limit of the local mass depends only on the asymptotic geometry. It is thus independent of a particular matter model representing the singular source. It actually follows directly from the formula \cref{eq:locmassuptorout} for the local mass of the matter distribution with a regular origin inside a given radius $r_\oix$. For radius $r_\oix=r_+$, which belongs to the asymptotic region, the result obviously depends only on the asymptotic geometry. For conical AdS asymptotics \cref{eq:RprofRPP}, we obtain
\begin{equation}
    \hat\locmass_+ = 1-R'_+ \quad\to\quad \hat\locmass_\oi = 1-c_\ai\;.
\end{equation}
Of course, the formula  \cref{eq:locmassuptorout} for the local mass holds only for the regular origin. The limit ${R_*\to 0}$ generalizes it to a singular case.

Notice that the same argument does not hold for the Killing mass. Formula \cref{eq:KVmassdomDregorg} (or its generalization analogous to \cref{eq:MKVshellsplit}, including contributions of shells) contains expressions evaluated not only at $r_+$, but also terms depending on the particular matter model ($N_\oi$ term and the shell terms).

Our discussion of various matter models demonstrates the existence of various plausible matter models representing a point-like source. It confirms the general result for the local mass $\hat\locmass_\oi$ and shows the results for the Killing mass, which depend on the choice of the model.


\subsection{Dust ball in vacuum}

We start with the simplest model of a static dust ball with a regular origin surrounded by a vacuum, without a cosmological constant. The surface of the sphere will be at radial coordinate $r=r_-$. First, we assume an arbitrary density of the dust $\eps_\dst(r)$ corresponding to the metric function $R(r)$, cf.~\cref{eq:epsdustR}. We do not need any details of this distribution except for the boundary values $R_*=R(r_-)$ and $R'_-=R'(r_-)$. The lapse inside the sphere is constant, $N=N_\oi$. The outer solution is given by the asymptotic metric \cref{eq:MinkConeGeom} parametrized by $N_\ai$ and $c_\ai$, with the surface of the sphere at $r=r_+$. Since we employed the freedom in choosing the origin of the coordinate $r$ for both inner and outer solutions separately, we cannot automatically assume that $r$ of the inner and outer solutions is continuous on the surface of the sphere. Inner and outer values of $r$ are thus $r_-$ and $r_+$. Coordinate $r$ could be changed to become continuous by shifting it by a constant in the outer region, but it would just introduce another parameter equivalent to the jump ${[r]=r_+-r_-}$.

In the ${\Lambda=0}$ case, we can require that there be no massive shell on the surface of the sphere. We thus have conditions ${[N]=[R]=0}$ and ${[N']=[R']=0}$, namely:
\begin{equation}
    N_\oi = N_\ai\;,\quad R_*=c_\ai r_+\;,\quad R'_-= c_\ai\;.
\end{equation}
Parameters of the inner solution $N_\oi$, $R_*$, $R'_-$ thus uniquely determine the parameters of the outer asymptotic geometry $N_\ai$, $c_\ai$, and the position $r_+$ of the surface with respect to the asymptotic geometry. Moreover, the lapse can be adjusted by normalizing the static Killing vector, and in this case, the natural choice with respect to both the inner and outer solutions is $N=1$.

The rescaled local mass up to a radius $r_\oix$ in the vacuum domain is given by \cref{eq:locmassuptorout},
\begin{equation}
    \hat\locmass
    = 1-c_\ai\;,
\end{equation}
and it does not depend on the choice of the boundary $r_\oix$ outside the sphere -- the whole mass is concentrated in the sphere.
The conicity $c_\ai$ is the only relevant parameter of the asymptotic geometry, and we see that it is given by the local mass. 

It also provides a nontrivial relation \cref{eq:locmassBTZmass} between the local mass and the BTZ parameter \cref{eq:BTZconicityL0}, namely ${1-\mBTZ = (1-\hat\locmass)^2}$. It transforms the positive/negative local mass $\hat\locmass$ to the positive/negative BTZ parameter $\mBTZ$, and both $\hat\locmass$ and $\mBTZ$ are restricted by the maximal value $\hat\locmass,\mBTZ<1$.

For homogeneous dust with constant density ${\eps_\dst=\eps_\oi}$ we have 
\begin{equation}
    R_*=\frac{1}{\sqrt{\kap\eps_\oi}}\sin\bigl(\sqrt{\kap\eps_\oi}\,r_-\bigr)\;,\quad
    R'_-=\cos\bigl(\sqrt{\kap\eps_\oi}\,r_-\bigr)\;.
\end{equation}
Therefore,
\begin{equation}
    c_\ai =\sqrt{1-\kap\eps_\oi R_*^2}\;,\quad
    r_+=\frac{R_*}{\sqrt{1-\kap\eps_\oi R_*^2}}\;.
\end{equation}
For the BTZ mass parameter, we obtain
\begin{equation}
    \mBTZ =\kap\eps_\oi R_*^2\;.
\end{equation}
We thus expressed the asymptotic characteristic of the geometry in terms of the dust density $\eps_\oi$ and the circumference radius $R_*$ of the sphere. 

Geometrically, the asymptotic conical geometry is glued to a spherical cap at radius $R_\oi=\frac1{\sqrt{\kap\eps_\oi}}$. The homogeneous spherical cap can be substituted with an arbitrary smooth tip. The only relevant quantities for the asymptotic geometry are the radius $R_*$ at the surface of the dust object and the conicity $c_\ai=R'_-$ given by the ratio of the circumference to the radial distance at the surface.


\subsection{Dust ball in AdS}
\label{ssec:DustAdS}

We are primarily interested in AdS asymptotics. Unfortunately, we proved in \cref{ssec:dust} that there is no static dust solution in the presence of the cosmological constant. Therefore, we consider a spherical bubble of spacetime with a vanishing cosmological constant filled with dust, surrounded by an AdS vacuum. In this case, the surface of the bubble must contain a thin shell $\Sigma$ that balances the pressure of the dark energy outside. We characterize the shell by its linear energy density $\eTS$ and angular tension $\pTS$. First, we do not assume any particular equation of state relating $\eTS$ and $\pTS$; later, we employ the Nambu-Goto equation of state \cref{eq:shellCS} and the radiation case \cref{eq:shellrad}.

As above, the dust solution determines $R_*$ and $R'_-$. The shell is phenomenologically described by dimensionless constants $\ehTS=\kap\eTS\ell$ and $\phTS=\kap\pTS\ell$.

\subsubsection*{Outer solution}

The outer asymptotic solution is now given by \cref{eq:asympAdSexp}. The junction conditions $[N]=0$, $[R]=0$  give 
\begin{equation}\label{eq:dustJcondval}
\begin{aligned}
    R_* &= R_\ai \frac12\Bigl(\exp\bigl({\tfrac{r_+}{\ell}}\bigr) \mp \exp\bigl({-\tfrac{r_+}{\ell}}\bigr)\Bigr)
       \;,\\
    N_\oi &= N_\ai \frac12\Bigl(\exp\bigl({\tfrac{r_+}{\ell}}\bigr) \pm \exp\bigl({-\tfrac{r_+}{\ell}}\bigr)\Bigr)
       \;,
\end{aligned}    
\end{equation}
and conditions \cref{eq:shellepsp} yield
\begin{equation}\label{eq:dustJcondder}
\begin{aligned}
    R'_-\ell-\ehTS R_* &= R_\ai \frac12\Bigl(\exp\bigl({\tfrac{r_+}{\ell}}\bigr) \pm \exp\bigl({-\tfrac{r_+}{\ell}}\bigr)\Bigr)
       \;,\\
    \phTS N_\oi &= N_\ai \frac12\Bigl(\exp\bigl({\tfrac{r_+}{\ell}}\bigr) \mp \exp\bigl({-\tfrac{r_+}{\ell}}\bigr)\Bigr)
       \;.
\end{aligned}    
\end{equation}
Dividing the proper relations in \cref{eq:dustJcondval} and \cref{eq:dustJcondder} we obtain 
\begin{equation}
    \frac{R_\ai}{N_\ai} 
      = \frac{ R'_-\ell-\ehTS R_*}{N_\oi} 
      = \frac{1}{\phTS}\frac{R_*}{N_\oi}\;,
\end{equation}
which provides the constraint
\begin{equation}\label{en:shellstbal}
   \frac{R'_-\ell}{R_*} = \ehTS + \frac{1}{\phTS}
\end{equation}
between the parameters of the inner dust solution and the shell. Further elementary manipulation of \cref{eq:dustJcondval} and \cref{eq:dustJcondder} yields
\begin{gather}
   \pm \frac{R_\ai^2}{R_*^2} = \frac{1}{\phTS^2}-1\;,\label{eq:dustAdSRa}\\
   \pm \exp\frac{2r_+}{\ell} = \frac{1+\phTS}{1-\phTS}\;.\label{eq:dustAdSrplus}
\end{gather}

We can make a couple of observations: We cannot avoid the shell in this case. Nay, the shell must have non-vanishing tension $\phTS$ to compensate for the jump in radial pressure between the dust inside and the dark energy outside. The value of $\phTS^2$ decides which asymptotic case applies. For $\phTS^2<1$ we have the upper sign in the equations, i.e., the conical asymptotic. For $\phTS^2>1$, we have the black hole asymptotic. The conditions \cref{eq:dustAdSRa} and \cref{eq:dustAdSrplus} can then be rewritten
\begin{equation}\label{eq:dustconpars}
   c_\ai = \frac{R_*}{\ell}\sqrt{\frac{1}{\phTS^2}-1}   \;,\quad
   \tanh\frac{r_+}{\ell} = \phTS
\end{equation}
for the conical asymptotics, and
\begin{equation}\label{eq:dustBHpars}
   R_\hor = R_*\sqrt{1-\frac{1}{\phTS^2}}   \;,\quad
   \coth\frac{r_+}{\ell} = \phTS 
\end{equation}
for the black hole asymptotics.

\subsubsection*{Inner solution}

Assuming a homogeneous dust with a positive energy density, the inner geometry is part of a sphere of radius $R_\oi=\frac{1}{\sqrt{\kap\eps_\oi}}$ from the origin up to a latitudinal angle ${\tht_-\equiv\frac{r_-}{R_\oi}}$, where
\begin{equation}\label{eq:dustsphsurf}
    R_\oi \sin\tht_- = R_*\;,\quad \cos\tht_-=R'_-\;.
\end{equation}
It allows relating the parameters of the inner solution $R_\oi$, $\tht_-$, $N_\oi$ to the parameters of the outer solution $R_\ai$, $r_+$, $N_\ai$ using \cref{eq:dustAdSRa}, \cref{eq:dustAdSrplus}, once the equation of state of the shell is specified.

\subsubsection*{Thin shell}

Since the shell tension $\phTS$ must be nonvanishing, we cannot employ the thin shell formed by incoherent dust particles. 

For the Nambu-Goto shell \cref{eq:shellCS}, substituting ${\ehTS=-\phTS}$ into \cref{en:shellstbal} and employing \cref{eq:dustconpars} or \cref{eq:dustBHpars}, respectively, and \cref{eq:dustsphsurf},  we obtain 
\begin{equation}
    \cos\tht_-=
        \frac{c_\ai}{\cosh\frac{r_+}{\ell}} 
\end{equation}
for the conical asymptotics, and
\begin{equation}
    \cos\tht_-=
        -\frac{R_\hor}{\ell\sinh\frac{r_+}{\ell}}        
\end{equation}
for the black hole asymptotics. We can observe that for the conical asymptotic, the outer solution is glued to a smaller part of the sphere, ${\tht_- < \frac\pi2}$; for the black hole asymptotic, it is glued to a part of the sphere larger than a hemisphere, ${\tht_- > \frac\pi2}$.

Analogously, for the trace-free stress-energy tensor of the shell, \cref{eq:shellrad}, one gets 
\begin{equation}
    \cos\tht_-=
        c_\ai\frac{\cosh\frac{2r_+}{\ell}}{\cosh\frac{r_+}{\ell}} 
\end{equation}
for the conical asymptotics, and
\begin{equation}
    \cos\tht_-=
        \frac{R_\hor}{\ell}\frac{\cosh\frac{2r_+}{\ell}}{\sinh\frac{r_+}{\ell}} 
\end{equation}
for the black hole asymptotics. In this case, the geometrical interpretation is not so obvious.

\subsubsection*{Mass}

\begin{figure}
    \centering
    \includegraphics[scale = .4]{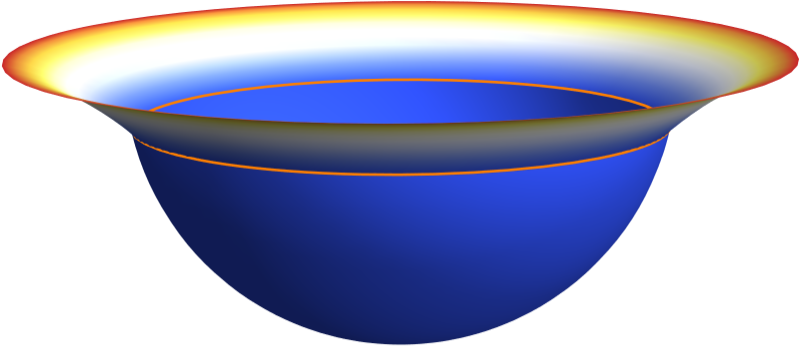}
    \caption{Embedding of the spatial section of the asymptotic conical AdS spacetime joined to the dust ball (disk in two spatial dimensions) with the geometry of a homogeneous spherical cap. The cosmological pressure of the surrounding spacetime is compensated by the thin Nambu-Goto shell (the orange circle). The color indicates a lapse, which is constant across the dust ball and grows with increasing radius outside it. The embedding region is maximal; more cannot be embedded into $\mathbb{E}^3$. Of course, the real asymptotic geometry is a hyperbolic one that extends up to infinity. Interestingly, the boundary of the embeddable region is exactly the circle with vanishing local mass inside it.}
    \label{fig:dust_con}
\end{figure}

\begin{figure}
    \centering
    \includegraphics[scale = .4]{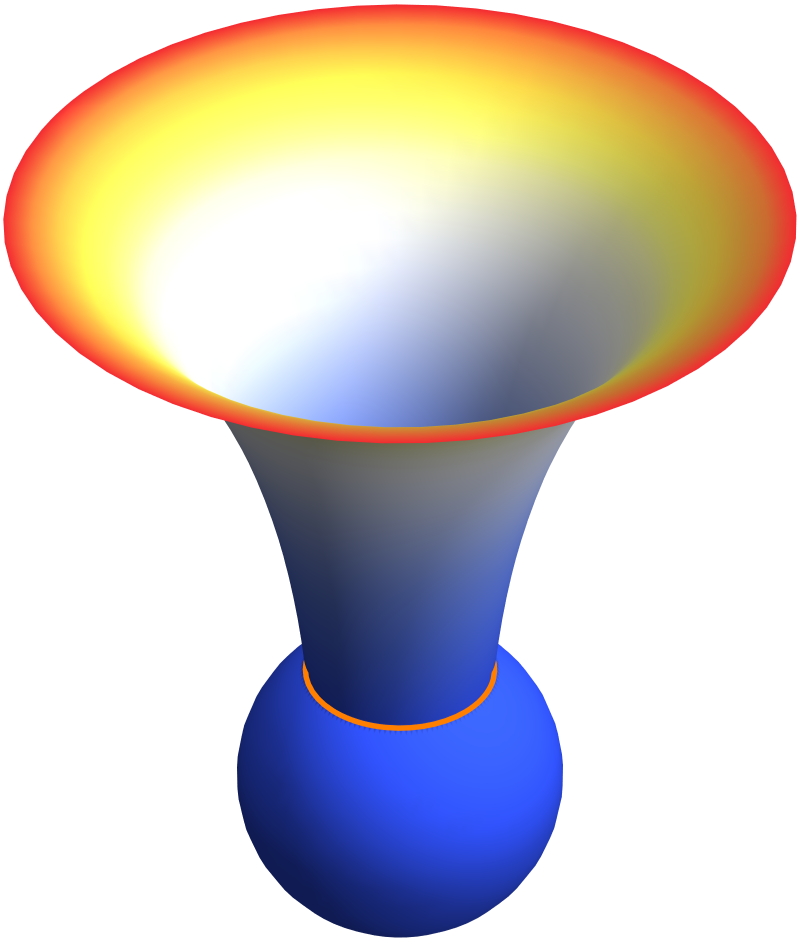}
    \caption{Embedding of the spatial geometry of the dust ball (disk) surrounded by the black-hole AdS asymptotic. The geometry of the dust ball is a part of a homogeneous sphere larger than a hemisphere. The color again encodes lapse, notably constant along the dust ball. The surface of the ball supports the thin shell (orange circle) that compensates for the cosmological pressure. The embedding region is again maximal and corresponds to a domain where the dust energy is exactly compensated by the dark-energy contribution.}
    \label{fig:dust_BH}
\end{figure}

Let us now discuss the local mass contained in a sphere of radius $r_\oix$ for the case of conical asymptotics. It is now justified to use the formula  \cref{eq:locmassuptorout}, since we have a model with a regular origin, and the formula for the local mass works even in the presence of the thin shell inside the studied region. For ${r_\oix>r_+}$, we thus have
\begin{equation}
    \hat\locmass^{r_\oix} = 1-c_\ai \cosh\frac{r_\oix}{\ell}\;.
\end{equation}
Of course, it decreases with growing $r_\oix$ since the dark energy in the asymptotic region. If we subtract the contribution of the dark energy above the dust sphere and shell, we obtain 
\begin{equation}\label{eq:locmassdustcon}
    \hat\locmass_+ = 1-c_\ai \cosh\frac{r_+}{\ell}\;.
\end{equation}
Taking the limit ${r_+\to 0}$ with fixed parameter $c_\ai$ of the asymptotic geometry, we obtain that the size of the ``curved tip'' decreases ${R_*=c_\ai\sinh\frac{r_+}{\ell}\to 0}$, but the mass hidden inside the tip is finite and, in general, non-trivial
\begin{equation}\label{eq:limitlocmassdustcon}
    \hat\locmass_+ \to \hat\locmass_\oi=1-c_\ai\;.
\end{equation}
For the homogeneous dust and the Nambu-Goto shell, we can find the limiting behavior of the parameters of the inner solution
\begin{equation}\label{eq:limitthtdustcon}
    \cos\tht_- \to c_\ai\;,
\end{equation}
and
\begin{equation}\label{eq:paramslimitthtdustcon}
   R_\oi \to 0\;,\quad
   \kap\eps_\oi \approx \frac{1}{r_+^2}\Bigl(\frac{1}{c_\ai^2}-1\Bigr)+\mathcal{O}(r_+^{\,0})\;.
\end{equation}
The radius of the spherical cap approaches zero, but the energy density increases in such a way that the mass of the dust remains finite, \cref{eq:limitlocmassdustcon}. The spherical cap is cut at the latitudinal angle $\tht_-$ given by the asymptotic conicity $c_\ai$, cf.~\cref{eq:limitthtdustcon}.

We should stress that the limiting mass \cref{eq:limitlocmassdustcon} of the dust object (including the shell) is determined solely by the asymptotic conical geometry (it depends only on $c_\ai$) and does not depend on a particular dust distribution inside the spherical region ${r<r_-}$. It is thus a rather robust justification of our earlier claim \cref{eq:locmasscontip} that the conical geometry \cref{eq:rot_sym_R} considered up to the semi-regular origin ${r=0}$ corresponds to the point-like particle of local mass \cref{eq:limitlocmassdustcon}.

For completeness, we also study the Killing mass. Due to the presence of the shell, we have to use formula \cref{eq:MKVshellsplit} to estimate the mass contained in the sphere larger than the dust and shell. Setting $r_\iix=0$ and $r_\oix=r_+$ in \cref{eq:MKVshellsplit}, we get the Killing mass of the dust and the shell
\begin{equation}
    \hat\MKV_+ =  N_\ai \cosh\frac{r_+}{\ell}\Bigl(1-c_\ai \cosh\frac{r_+}{\ell}\Bigr)\;.
\end{equation}

In the limit $r_+\to 0$, we again obtain
\begin{equation}
    \label{eq:KillingMassDustConus}
    \hat\MKV_+ \to  \hat\MKV_\oi = N_\ai ( 1-c_\ai ) = N_\ai\hat\locmass_\oi\;.
\end{equation}

\subsubsection*{Vacuum bubble}

Our general discussion also covers the special case of a spherical bubble with no matter surrounded by a thin shell. It corresponds to $\eps_\oi=0$. The inner geometry changes to a flat ball (disk) of radius $R_*$, given by the metric function $R(r)=r$, which means $R'_-=1$. For the Nambu-Goto shell and conical asymptotics, it implies the restrictions
\begin{equation}
    c_\ai = \cosh\frac{r_+}{\ell} \;.
\end{equation}
The asymptotic conicity uniquely determines the size of the bubble. We cannot thus make a limit of a small bubble\footnote{A small bubble means ${r_+\to 0}$. Indeed, the circumference radius of the bubble is ${R_* = c_\ai\ell\sinh\frac{r_+}{\ell}}\to 0$.} with a fixed asymptotic conicity $c_\ai$. Making the bubble small implies ${c_\ai\to1}$, which sends the mass of the bubble with the shell to zero. For black hole asymptotics, the corresponding constraint reads:
\begin{equation}
    R_\hor= -\ell\sinh\frac{r_+}{\ell}
\end{equation}
which cannot be satisfied if we require the bubble at ${r_+>0}$, i.e., above the horizon.

Assuming the shell formed by the radiative matter, we obtain constraints
\begin{equation}
    c_\ai = \frac{\cosh\frac{r_+}{\ell}}{\cosh\frac{2r_+}{\ell}} 
\end{equation}
for the conical asymptotics, and
\begin{equation}
    \frac{R_\hor}{\ell} =\frac{\sinh\frac{r_+}{\ell}}{\cosh\frac{2r_+}{\ell}}
\end{equation}
for the black hole asymptotics. Both asymptotics are possible in this case. However, neither case provides a non-trivial limit of a small bubble.


\subsection{Fluid ball in AdS}
\label{ssec:FluidAdS}

Let us now consider a spherical object composed of the incompressible fluid (cf.~\cref{ssec:fluid}) in an AdS background. We assume the same cosmological term $\Lambda=-\frac{1}{\ell^2}$ throughout the entire spacetime, and we do not allow for a thin shell on the surface of the fluid.

\subsubsection{Inner solution}

The rescaled fluid energy density~$\hat\eps_\fld$ and pressure~$\hat{p}_\fld$ differ from the total quantities $\hat\eps$, $\hat{p}$ by the cosmological (dark energy) term,
\begin{equation}\label{eq:modEpsP}
    \hat\eps
    = -1 + \hat\eps_\fld\;,\quad
    \hat{p}
    = 1 + \hat{p}_\fld\;,
\end{equation}
cf.~\cref{eq:rescaledepsp}, \cref{eq:Lambdaepsp}, and footnote \ref{fnt:fluidLambda}.
We denote their values at the origin as $\hat\eps_\oi$ and $\hat{p}_\oi$, respectively:\footnote{Notice, that $\eps_\oi$ and $p_\oi$ in \cref{ssec:fluid} included the cosmological term. Here, we exclude it from the definition of $\hat\eps_\oi$ and $\hat{p}_\oi$.} 
\begin{equation}
    \label{eq:modEpsPorg}
    \hat\eps_\oi = \hat\eps_\fld\;, \quad  \hat{p}_\oi = \hat{p}_\fld|_{r=0}\;.
\end{equation}
Recall that the energy density $\hat\eps_\fld$ is constant inside the object.

The fluid ball extends up to radial coordinate $r_-$ with the circumference radius $R_*$. We do not allow a thin shell on the surface. Thus, we expect that the metric function $R'$ is continuous there, and we denote its value $R'_*$,
\begin{equation}\label{eq:fluidjuctRm}
  R(r_-)=R_* \;,\quad R'(r_-) = R'_*\;.
\end{equation}

For general $\hat{\eps}_\oi$ and $\hat{p}_\oi$, the fluid pressure $\hat{p}_\fld$ vanishes at a certain radius. To avoid a jump in the pressure on the surface of the ball (which would require a thin shell), we cut the fluid solution exactly when $\hat{p}_\fld(r_-) = 0$. Substituting \cref{eq:modEpsP} into \cref{eq:fluid_RderR} and \cref{eq:fluidpressureR} and taking into account conditions on the surface, we find
\begin{equation}\label{eq:fluidsurfcond}
\begin{gathered}
    R'_*{}^2 = 1-(\hat\eps_\oi{-}1)\frac{R_*^2}{\ell^2}\;,\quad
    1 = \frac{(\hat{p}_\oi+1) R'_*}{1+\frac{\hat{p}_\oi+1}{\hat\eps_\oi-1}(1-R'_*)}\;.
\end{gathered}
\end{equation}
We can solve for the central energy density and pressure in terms of the quantities $R_*$, $R'_*$ defined on the surface:
\begin{equation}\label{eq:fluidballcentrepsp}
\begin{gathered}
    \hat{\eps}_\oi = 1+\frac{1-R'_*{}^2}{\frac{R_*^2}{\ell^2}}\;,\\
    \hat{p}_\oi = \frac{1-R'_*{}^2+\frac{R_*^2}{\ell^2}}{R'_*(1+R'_*)-\frac{R_*^2}{\ell^2}}\;.
\end{gathered}
\end{equation}
Inverse relations are
\begin{equation}\label{eq:fluidballcentrRRps}
\begin{gathered}
    \frac{R_*^2}{\ell^2} = \frac{1}{\hat\eps_\oi}\frac{\hat{p}_\oi}{\hat{p}_\oi+1}
      \Bigl(1+\frac{\hat\eps_\oi+\hat{p}_\oi}{\hat\eps_\oi(\hat{p}_\oi+1)}\Bigr)\;,\\
    R'_* = \frac{\hat\eps_\oi+\hat{p}_\oi}{\hat\eps_\oi(\hat{p}_\oi+1)}\;.
\end{gathered}
\end{equation}%
Equations \cref{eq:fluid_Rlapse} and \cref{eq:fluid_Rlapseder} provide the corresponding relations for the lapse.

If the central total energy density $\eps$ is positive (i.e., ${\hat\eps_\oi>1}$), the spatial geometry of the inner solution is again that of a section of a homogeneous ball of radius 
\begin{equation}\label{eq:fluidballRo}
R_\oi=\frac{1}{\sqrt{\kap\eps}}=\frac\ell{\sqrt{\hat\eps_\oi-1}}\;. 
\end{equation}
Introducing the latitudinal angle ${\tht=\frac{r}{R_\oi}}$, we have
\begin{equation}
    R = R_\oi \sin\tht\;,\quad R' = \cos\tht\;.
\end{equation}
We see that on the surface it must hold ${R_*<R_\oi}$ and ${R'_*<1}$. The surface of the fluid is given by \cref{eq:fluidjuctRm},
 \begin{equation}\label{eq:fluidsurfrm}
     \cos\tht_- = R'_*
       =\frac{\hat\eps_\oi+\hat{p}_\oi}{\hat\eps_\oi(\hat{p}_\oi+1)}\;.
 \end{equation}%
Assuming ${R'_*>0}$, we get that the homogeneous ball must be cut at ${\tht_-<\frac\pi2}$, i.e., inside the hemisphere of the origin. 

The central pressure need not be positive, but we usually require ${p_\fld>0}$ to ensure reasonable physical behavior. The condition for the positivity of the central pressure $\hat{p}_\oi$ is
\begin{equation}
    R'_*(1+R'_*) - \frac{R_*^2}{\ell^2} \ge 0\;,
\end{equation}
cf.~\cref{eq:fluidballcentrepsp}. Eliminating $R'_*$ using, e.g., \cref{eq:fluidsurfcond} we find
\begin{equation}    \label{eq:posPressure}
    R_* \le \ell\frac{\sqrt{\hat\eps_\oi+1}}{\hat\eps_\oi}\;.
\end{equation}
Such $R_*$ automatically satisfies ${R_*<R_\oi}$ and ${R'_*<1}$.

For ${\hat\eps_\oi<1}$, the spatial geometry of the inner solution would be hyperbolic. The discussion would be analogical, and we skip the details here.

In summary, the inner solution is parameterized by the central fluid energy density $\hat\eps_\oi$ and pressure $\hat{p}_\oi$. They already determine the position of the fluid surface by \cref{eq:fluidballcentrRRps} and \cref{eq:fluidsurfrm}. They can be reexpressed in terms of the parameters of the surface $R_*$ and $R'_*$ using \cref{eq:fluidballcentrepsp}.

\subsubsection{Outer solution}

Now we want to join the inner solution describing the fluid ball to the outer solution of empty AdS spacetime discussed in \cref{sec:AsymptGeom}. We are interested in the situation in which the surface of the ball does not contain any singular matter distribution. The condition for the absence of the shell \cref{eq:shellepsp} requires that $R(r)$ and $N(r)$ are continuous and have continuous first derivatives, $[R']=0$, $[N']=0$. The equation \cref{eq:fluidEq} for the radial pressure then implies that the pressure (including the dark energy contribution) must be continuous, $[p]=0$. Therefore, to avoid a thin shell, the pressure $p_\fld$ of the fluid must vanish at the surface of the fluid object, as we required above. 

The outer solution \cref{eq:asympAdSexp} starts at the radial coordinate $r_+$. Since, as before, we set the origin of the coordinate $r$ separately in the inner and outer domains, the radial coordinate is not continuous at the junction between the inner and outer solutions, ${r_- \neq r_+}$. However, the outer solution satisfies
\begin{equation}\label{eq:fluidjuctRp}
  R(r_+)=R_* \;,\quad R'(r_+) = R'_*\;,
\end{equation}
and the radius $r_+$ is given by the junction conditions relating these to \cref{eq:fluidjuctRm}.

Before we compare \cref{eq:fluidjuctRm} to \cref{eq:fluidjuctRp} and derive a relation between the parameters of the inner and outer spatial geometries, let us look at the lapse. In agreement with the isotropy condition \cref{eq:fluidisotropy}, the relations \cref{eq:asympAdSexp} imply $N'=\frac{N_\ai}{R_\ai}\frac{R}{\ell}$. Together with \cref{eq:fluid_Rlapseder} we obtain
\begin{equation}\label{eq:fluidlapserel}
    \frac{N_\ai}{N_\oi}=\frac{R_\ai}{\ell}(\hat{p}_\oi+1)\;.
\end{equation}
It relates the normalizations of the lapse of the inner and outer solution once we express $R_\ai$ in terms of $\hat\eps_\oi$ and $\hat{p}_\oi.$

\subsubsection{Conical AdS asymptotics}

Let us now investigate the case when the outer solution is a conical AdS solution \cref{eq:asympAdScon}. At the junction surface
\begin{equation}\label{eq:fluidRsRpscon}
    R_*=c_\ai\ell\sinh\frac{r_+}{\ell}\;,\quad R'_* = c_\ai\cosh\frac{r_+}{\ell}\;.
\end{equation}
It justifies the assumption $R'_*>0$ we made above. From the continuity of both $R$ and $R'$, it follows
\begin{gather}
  c_\ai^2 = R'_*{}^2-\frac{R_*^2}{\ell^2}= 
      \frac{\hat\eps_\oi-\hat{p}_\oi^2}{\hat\eps_\oi(\hat{p}_\oi+1)^2}\;,\label{eq:fluidconca}\\
  \tanh{\frac{r_+}{\ell}}=
      \frac{R_*}{R'_*\ell} 
      =\frac{\sqrt{\hat{p}_\oi(\hat\eps_\oi\hat{p}_\oi+2\hat\eps_\oi+\hat{p}_\oi)}}{\hat\eps_\oi+\hat{p}_\oi}
      \;,\label{eq:fluidconrp}
\end{gather}
where we used expressions \cref{eq:fluidballcentrRRps}. The inverse relations yield substituting \cref{eq:fluidRsRpscon} into \cref{eq:fluidballcentrepsp},
\begin{gather}
    \hat\eps_\oi = \frac{1-c_\ai^2}{c_\ai^2\sinh^2\!\frac{r_+}{\ell}}\;,\label{eq:fluidcentepscon}\\    
    \hat{p}_\oi = \frac{1-c_\ai^2}{c_\ai^2 + c_\ai\cosh\frac{r_+}{\ell}}\;.\label{eq:fluidcentpcon}
\end{gather}

Recall that ${c_\ai>0}$. We see that for ${0<c_\ai<1}$, both the energy density and pressure are positive. The case ${c_\ai=1}$ describes an empty AdS with the regular origin, ${\hat\eps_\oi=0}$, ${\hat{p}_\oi=0}$. The case ${c_\ai>1}$ corresponds to the asymptotic excess angle. It implies negative energy density and central pressure.

Negative total central energy density, i.e., ${\hat\eps_\oi-1<0}$, requires a hyperbolic spatial geometry for the inner solution. In terms of asymptotic parameters, the hyperbolic inner geometry applies if
\begin{equation}
    c_\ai\cosh\frac{r_+}{\ell}>1\;,
\end{equation}
cf.~\cref{eq:fluidcentepscon}. Thus, for ${c_\ai>1}$, and even for ${c_\ai<1}$ and large $r_+$, the inner spatial geometry must be hyperbolic. However, for $c_\ai<1$ and a sufficiently small radius $r_+$, we always find a spherical inner solution discussed in detail above.

Let us assume the spherical spatial geometry, ${\hat\eps_\oi>1}$. Since ${\frac{R_*}{R'_*\ell}=\tanh{\frac{r_+}{\ell}}<1}$, the first equation in \cref{eq:fluidsurfcond} yields
\begin{equation}
    R_* < \frac{\ell}{\sqrt{\hat{\eps}_\oi}}\;
\end{equation}
Thus, for conical asymptotic AdS geometry and a given central energy density $\hat{\eps}_\oi>1$, the circumference radius $R_*$ of the fluid ball is bounded from above. This condition is a stronger restriction than \cref{eq:posPressure}. The central pressure for spherical inner geometry is thus always positive.

\begin{figure}
    \centering
    \includegraphics[width=\linewidth]{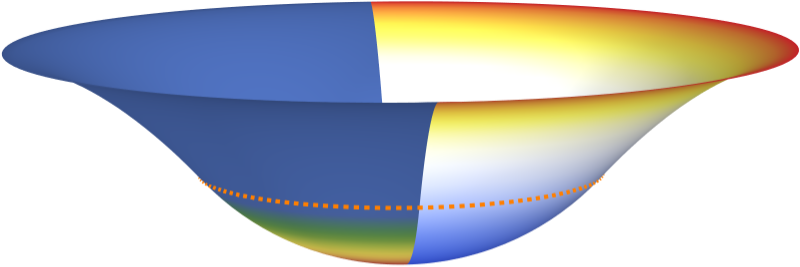}
    \caption{Embedding of the spatial section of the asymptotically conical AdS spacetime with the central region replaced by fluid. The orange dotted ring denotes the boundary between these two regions, but since the metric functions and their derivatives are continuous there, there is no notable kink in the embedding surface. The left side of the picture, with more saturated colors, encodes the pressure. It is maximal at the center of the fluid and decreases towards the boundary to the cosmological value, being constant outside the fluid. The right-hand side encodes the lapse. Unlike in the case with the dust, in this case, the lapse increases continuously from the center.}
    \label{fig:fluid_con}
\end{figure}

\subsubsection{Black hole asymptotics}

For the outer solution \cref{eq:asympAdSBH} with the black hole asymptotics, we have
\begin{equation}\label{eq:fluidRsRpsBH}
    R_*=R_\hor\cosh\frac{r_+}{\ell}\;,\quad R'_* = \frac{R_\hor}{\ell}\sinh\frac{r_+}{\ell}\;.
\end{equation}
The assumption ${R'_*>0}$ made above means that the surface of the fluid is above the black hole horizon.\footnote{The case when ${R'_*<0}$, i.e., ${r_+<0}$, could also be considered. It would describe situations such as a fluid ball with a small cavity (smaller than the ball's hemisphere) containing a black hole. In a sense, it is complementary to the solution for a small fluid ball with black hole asymptotics discussed in the text.} Using \cref{eq:fluidballcentrRRps}, we obtain
\begin{gather}
  R_\hor^2 = R_*^2-R'_*{}^2\ell^2= 
      \frac{\hat\eps_\oi-\hat{p}_\oi^2}{\hat\eps_\oi(\hat{p}_\oi+1)^2}\;,\\
  \coth{\frac{r_+}{\ell}}=
      \frac{R_*}{R'_*\ell} 
      =\frac{\sqrt{\hat{p}_\oi(\hat\eps_\oi\hat{p}_\oi+2\hat\eps_\oi+\hat{p}_\oi)}}{\hat\eps_\oi+\hat{p}_\oi}
      \;.
\end{gather}
The inverse relations are
\begin{equation}
    \begin{gathered}
        \hat\eps_\oi = \frac{\ell^2+R_\hor^2}{R_\hor^2\cosh^2\!\frac{r_+}{\ell}}\;,\\    
        \hat{p}_\oi = \frac{\ell^2+R_\hor^2}{R_\hor\ell\sinh\frac{r_+}{\ell}-R_\hor^2}\;.
    \end{gathered}
\end{equation}

The condition of the spherical inner spatial geometry, ${\hat\eps_\oi>1}$, in this case yields 
\begin{equation}
    \frac{R_\hor}{\ell}\sinh\frac{r_+}{\ell}<1\;.
\end{equation}
Restricting to this case, the upper bound for the circumference radius $R_*$ is given by the radius $R_\oi$ of the spherical geometry \cref{eq:fluidballRo}. The lower bound for $R_*$ follows from the relation $\frac{R_*}{R'_*\ell}=\coth{\frac{r_+}{\ell}}>1$. Together
\begin{equation}
    \frac{\ell}{\sqrt{\hat{\eps}_\oi}}< R_* < \frac{\ell}{\sqrt{\hat{\eps}_\oi-1}}\;
\end{equation}
For a given central energy density $\hat{\eps}_\oi>1$, a circumference ball's radius $R_*$ greater than $\frac{\ell}{\sqrt{\hat\eps_\oi}}$ requires the black-hole asymptotic AdS geometry. For $R_*$ breaking \cref{eq:posPressure}, we need negative central pressure.

\begin{figure}
    \centering
    \includegraphics[width=\linewidth]{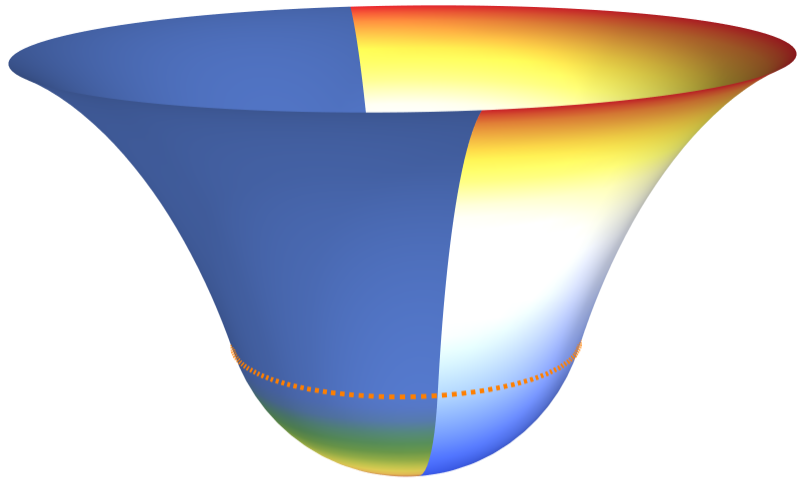}
    \caption{Embedding of the spatial section of the asymptotically black hole AdS spacetime with the central region replaced by fluid. The color coding is the same as in \cref{fig:fluid_con}. The difference from the conical case is not as prominent as it was for the dust solution, since the boundary is continuous with continuous first derivatives. But generally the figure is more elongated in agreement with the fact that it limits towards a black hole throat while it limits to the tip of the cone in \cref{fig:fluid_con}.}
    \label{fig:fluid_BH}
\end{figure}

\subsubsection{Mass}

Let us fix the outer solution with the conical asymptotics, with a cosmological scale $\ell$, conicity $c_\ai$, and the lapse normalization $N_\ai$. Let the inner solution be the fluid ball with spherical spatial geometry and a surface at the circumference radius $R_*$. We can always find central energy density $\eps_\oi$ and pressure $\hat{p}_\oi$ to have a viable fluid solution below $R_*$. The results are
\begin{equation}\label{eq:floofa}
    \begin{gathered}
      \hat{\eps}_\oi = \frac{\ell^2}{R_*^2}\bigl(1-c_\ai^2\bigr)\;,\\
      \hat{p}_\oi = \frac{1-c_\ai^2}{c_\ai^2+\sqrt{c_\ai^2+\frac{R_*^2}{\ell^2}}}\;,\\
      N_\oi = \frac{N_\ai}{c_\ai} \frac{c_\ai^2+\sqrt{c_\ai^2+\frac{R_*^2}{\ell^2}}}{1+\sqrt{c_\ai^2+\frac{R_*^2}{\ell^2}}}\;,
    \end{gathered}
\end{equation}
cf.~\cref{eq:fluidconca}, \cref{eq:fluidcentpcon}, \cref{eq:fluidRsRpscon}, and \cref{eq:fluidlapserel}.

We can now compute the local mass up to the radius $R_*$, cf.~\cref{eq:locmassuptorout},
\begin{equation}
    \hat\locmass_* 
    = 1-R'_* 
    = 1-\sqrt{c_\ai^2+\frac{R_*^2}{\ell^2}}\;.
\end{equation}
In the limit $R_*\to 0$, the local mass behaves 
\begin{equation}
    \hat\locmass_* \to \hat\locmass_\oi = 1-c_\ai\;,
\end{equation}
the same as for the dust bubble \cref{eq:limitlocmassdustcon}. The fluid energy density diverges similarly to \cref{eq:paramslimitthtdustcon}, for the other parameters, it holds
\begin{equation}
    \begin{gathered}
      \hat{p}_\oi \to \frac{1-c_\ai}{c_\ai}\;,\\
      N_\oi \to N_\ai\;.
    \end{gathered}
\end{equation}

Computing the Killing mass using \cref{eq:KVmassdomDregorg}, the result for the fluid ball of radius $R_*$ is
\begin{equation}
  \hat\MKV_* = \frac{N_\ai}{c_\ai}
    \Biggl((1-c_\ai^2)\frac{\sqrt{c_\ai^2+\frac{R_*^2}{\ell^2}}}{1+\sqrt{c_\ai^2+\frac{R_*^2}{\ell^2}}}
    -\frac12\frac{R_*^2}{\ell^2}\Biggr)\;.
\end{equation}
The limit $R_*\to0$ again reveals \cref{eq:KillingMassDustConus},
\begin{equation}\label{eq:KillingMassFluid}
  \hat\MKV_* \to \hat\MKV_\oi = N_\ai (1-c_\ai)\;.
\end{equation}

We can conclude that the conical AdS geometry, smoothed near the origin by the ball of incompressible fluid, simulates, in the limit of a small ball, a point particle of the local mass $\hat\locmass_\oi=1-c_\ai$. It is the same result as for the dust ball, derived previously.


\subsection{Dark energy bubble in AdS}

Let us now consider the case where the inner region is formed by a regular vacuum spacetime with a generic cosmological constant $\Lambda'$ different from the cosmological constant $\Lambda=-\frac{1}{\ell^2}$ of the outer region. We can view this as a bubble of dark energy (of a generic sign of energy density) confined by a thin shell within the cosmological AdS background.

\subsubsection{Inner solution}

The inner solution is a regular $\Lambda'$-vacuum spacetime discussed in \cref{sec:AsymptGeom}. We denote the energy density and pressure corresponding to the inner cosmological constant as 
\begin{equation}
  \kap\eps_\oi=-\kap p_\oi=\Lambda'\;,
\end{equation}
and their dimensionless version is obtained by rescaling by the asymptotic cosmological scale $\ell$,
\begin{equation}
  \hat\eps_\oi = \kap\eps_\oi \ell^2\;,\quad\hat{p}_\oi= \kap p_\oi \ell^2\;.
\end{equation}
Notice that ${1+\hat\eps_\oi=(\Lambda'-\Lambda)\ell^2}$ characterizes the energy density added to the asymptotic cosmological \mbox{$\Lambda$-contribution}. If we introduce the inner cosmological scale $\ell'$ by ${\Lambda'=\pm\frac1{\ell'^2}}$, the dimensionless energy density $\hat\eps_\oi$ gives the ratio of both scales, ${\hat\eps_\oi=\pm\frac{\ell^2}{\ell'^2}}$. Thus, $\hat\eps_\oi=-1$ represents no bubble and $\hat\eps_\oi=0$ a bubble with no total dark energy (cf.\ vacuum bubble in \cref{ssec:DustAdS}).

The requirement of the regular origin inside the bubble fixes the conicity parameter ${c_\ai=1}$ in the solutions discussed in \cref{sec:AsymptGeom}. We also rename the lapse normalization $N_\ai$ to $N_\oi$. We thus consider solutions \cref{eq:MinkConeGeom}, \cref{eq:asympAdScon}, or \cref{eq:asympdScon} for $\eps_\oi$ being zero, negative, or positive, respectively. The metric can be written
\begin{equation}
    \label{eq:bubbleBTZform}
	\ts{g} = -(1-\kap\eps_\oi R^2)\dif T^2 
        + \frac{1}{1-\kap\eps_\oi R^2}\, \dif R^2 
        + R^2\,\dif \ph^2 \;.
\end{equation}

The inner solution is valid inside the ball (disk) of radius $r_-$. The values of the metric functions at the boundary of the bubble are
\begin{gather}
        R(r_-) = R_*\;,\quad
        R'(r_-) = \sqrt{1-\kap\eps_\oi R_*^2}\;,\\
        N(r_-) = N_\oi \sqrt{1-\kap\eps_\oi R_*^2}\;,\quad
        N'(r_-) = -\kap\eps_\oi N_\oi R_* \;.\notag
\end{gather}
Clearly, ${1>\kap\eps_\oi R_*^2}$ must hold. It restricts the case of positive dark energy density when the size $R_*$ of the bubble must be smaller than the radius ${R_\oi=\frac{1}{\sqrt{\kap\eps_\oi}}}$ given by energy density, i.e., ${R_*<R_\oi}$.

\subsubsection{Outer solution}

We consider that the bubble of dark energy is immersed in AdS asymptotics, with a cosmological scale~$\ell$. As before, we have to distinguish the conical and black hole asymptotics. 

For the conical case \cref{eq:asympAdScon}, the values of the metric functions of the conical geometry at the boundary of the bubble are
\begin{equation}
    \begin{gathered}
        R(r_+) = R_*\;,\quad
        R'(r_+) = \sqrt{c_\ai^2+\frac{R_*^2}{\ell^2}}\;,\\
        N(r_+) = \frac{N_\ai}{c_\ai} \sqrt{c_\ai^2+\frac{R_*^2}{\ell^2}}\;,\quad
        N'(r_+) = \frac{N_\ai}{c_\ai \ell^2} R_* \;.
    \end{gathered}
\end{equation}
Junction conditions require continuous $N$, which implies
\begin{equation}
    N_\ai = c_\ai N_\oi\sqrt{\frac{1-\hat\eps_\oi \frac{R_*^2}{\ell^2}}{c_\ai^2+\frac{R_*^2}{\ell^2}}}\;.
\end{equation}

For the black hole asymptotics, the metric functions of the black hole region at the boundary of the bubble take values
\begin{equation}
  \begin{gathered}
    R(r_+) = R_*\;,\quad
    R'(r_+) = \frac1\ell\sqrt{R_*^2-R_\hor^2}\;,\\
    N(r_+) = \frac{N_\ai}{R_\hor} \sqrt{R_*^2-R_\hor^2}\;,\quad
    N'(r_+) = \frac{N_\ai}{R_\hor \ell} R_* \;.
  \end{gathered}
\end{equation}
The junction condition for the lapse yields
\begin{equation}
    N_\ai = \frac{R_\hor}{\ell} N_\oi\sqrt{\frac{\ell^2-\hat\eps_\oi R_*^2}{R_*^2-R_\hor^2}}\;.
\end{equation}

The dark energy bubble must be confined by a thin shell. Its energy density $\eTS$ and pressure $\pTS$ are given by \cref{eq:shellepsp}. They should be related by a suitable equation of state.

\subsubsection{Nambu-Goto equation of state}

Let us assume that the thin shell follows the Nambu-Goto equation of state \cref{eq:shellCS}, which implies
\begin{equation}
    \frac{[R']}{R_*} = \frac{[N']}{N_*}\;.
\end{equation}

For the black-hole asymptotics, this relation requires
\begin{equation}\label{eq:bubbleCSBHasymp}
    R_\hor^2\sqrt{\ell^2-\hat\eps_\oi R_*^2} + \ell^2\sqrt{R_*^2-R_\hor^2} = 0\;.
\end{equation}
It cannot be fulfilled\footnote{We consider only the region above the black hole horizon where ${N>0}$, which means ${\sqrt{\ell^2-\hat\eps_\oi R_*^2}>0}$ and ${\sqrt{R_*^2-R_\hor^2}>0}$. We thus do not consider a bubble with a cavity that contains a real black hole with the black hole asymptotics behind the Einstein-Rosen bridge.} and the black hole asymptotics is thus ruled out.

For the conical asymptotics, the equation of state implies
\begin{equation}
    \sqrt{1-\hat\eps_\oi \frac{R_*^2}{\ell^2}} = \frac{1}{c_\ai^2}\sqrt{c_\ai^2+\frac{R_*^2}{\ell^2}}\;,
\end{equation}
which gives a simple relationship for the central and asymptotic lapse
\begin{equation}
    N_\oi = c_\ai N_\ai\;.
\end{equation}
After some manipulations, one can find the asymptotic conicity in terms of the central energy density and the bubble size
\begin{equation}\label{eq:bubbleCScai}
    c_\ai = \sqrt{\frac{1+\sqrt{1+4\frac{R_*^2}{\ell^2}\bigl(1-\hat\eps_\oi \frac{R_*^2}{\ell^2}\bigr)}}{2\bigl(1-\hat\eps_\oi \frac{R_*^2}{\ell^2}\bigr)}}\;.
\end{equation}
For a bubble satisfying the obligatory condition ${1>\kap\eps_\oi R_*^2}$ (a restriction on the bubble size $R_*$ for ${\eps_\oi>0}$), there exists always a solution. 

Next, we express various quantities in terms of the asymptotic conicity $c_\ai$ and the size $R_*$ of the bubble. The inverse relation \cref{eq:bubbleCScai} yields
\begin{equation}
    \hat\eps_\oi 
    = -\frac1{c_\ai^4}+\frac{\ell^2}{R_*^2}\Bigl(1-\frac1{c_\ai^2}\Bigr)\;.
\end{equation}
The linear energy density $\ehTS$ of the shell is
\begin{equation}
    \ehTS
    = \frac{\ell}{R_*}\Bigl(\frac1{c_\ai^2}-1\Bigr)\sqrt{c_\ai^2+\frac{R_*^2}{\ell^2}}\;.
\end{equation}
The local mass up to radius $r$ is given by \cref{eq:locmassuptorout}. The mass of the bubble, without the shell and including the shell, thus reads:
\begin{align}
    \hat\locmass_- &= 1- \frac1{c_\ai^2}\sqrt{c_\ai^2+\frac{R_*^2}{\ell^2}}\;,\\
    \hat\locmass_+ &= 1- \sqrt{c_\ai^2+\frac{R_*^2}{\ell^2}}\;.
\end{align}
As we discussed in \cref{sec:mass}, it holds ${\hat\locmass_+=\hat\locmass_- +\hat\locmass_\shl}$ with ${\hat\locmass_\shl= \frac{\ehTS}{\ell}R_*}$.

For sufficiently small conicity, the required energy density $\hat\eps_\oi$ of the bubble is negative. Counterintuitively, for conicity that is large enough, ${c_\ai^2>\frac12\Bigl(1+\sqrt{1+4\frac{R_*^2}{\ell^2}}\Bigr)}$ (which always corresponds to the excess angle), the required energy density $\hat\eps_\oi$ is positive. The energy density of the shell $\ehTS$, as well as its local mass $\hat\locmass_\shl$, is positive for $0<c_\ai<1$ and negative for $c_\ai>1$. The total local mass $\hat\locmass_+$ of the bubble and the shell is positive for $c_\ai^2+\frac{R_*^2}{\ell^2}<1$ and is composed of a negative contribution  $\hat\locmass_-$ from the bubble and a larger positive contribution $\hat\locmass_\shl$ from the shell. Surprisingly, the positive energy inside the bubble implies a large negative energy of the shell, yielding the total local mass $\hat\locmass_+$ to be negative.

The Killing mass of the bubble with the shell splits
\begin{equation}
\hat\MKV_+=\hat\MKV_- +\hat\MKV_\shl
\end{equation}
where
\begin{equation}
\begin{aligned}
    \hat\MKV_- &= 
      \frac12 \frac{N_\ai}{c_\ai}\Bigl(c_\ai^2-1-\frac{R_*^2}{c_\ai^2\ell^2}\Bigr)\;,\\
    \hat\MKV_\shl &= \frac{N_\ai}{c_\ai}(1-c_\ai^2)\Bigl(1+\frac{R_*^2}{c_\ai^2\ell^2}\Bigr)\;.
\end{aligned}
\end{equation}

Now we can perform the limit of a small bubble, ${R_*\to 0}$. We obtain
\begin{equation}
\begin{aligned}
    \hat\locmass_- &\to 1- \frac1{c_\ai}\;,\\
    \hat\locmass_\shl &\to \frac1{c_\ai}- c_\ai\;,\\
    \hat\locmass_+ &\to \hat\locmass_\oi=1- c_\ai\;.
\end{aligned}
\end{equation}
and
\begin{equation}
\begin{aligned}
    \hat\MKV_- &\to -\frac12 N_\ai\Bigl(\frac{1}{c_\ai}-c_\ai\Bigr)\;,\\
    \hat\MKV_\shl &\to N_\ai\Bigl(\frac{1}{c_\ai}-c_\ai\Bigr)\;,\\
    \hat\MKV_+ &\to \hat\MKV_\oi = \frac12 N_\ai\Bigl(\frac{1}{c_\ai}-c_\ai\Bigr)\;.
\end{aligned}
\end{equation}
We see that the asymptotic conicity ${0<c_\ai<1}$ corresponds to a positive mass $\hat\locmass_+$; however, the mass inside the bubble is negative and is compensated by a positive mass from the shell. Surprisingly, in this case, the Killing mass $\hat\MKV_\oi$ of the negligibly small bubble is not related to the local mass $\hat\locmass_\oi$ by the relation $\hat\MKV_\oi=N_\ai \hat\locmass_\oi$, cf.\ \cref{eq:KillingMassDustConus}, \cref{eq:KillingMassFluid}.

\subsubsection{Radiative equation of state}

Let us focus now on the shell satisfying the radiation equation of state \cref{eq:shellrad},
\begin{equation}
    \frac{[R']}{R} = -\frac{[N']}{N}\;.
\end{equation}
For the conical asymptotics, it implies the condition
\begin{equation}\label{eq:bubbleRadconasymp}
    \Bigr(c_\ai^2+2\frac{R_*^2}{\ell^2}\Bigr)\sqrt{1-\hat\eps_\oi \frac{R_*^2}{\ell^2}} = 
    \bigl(1-2\hat\eps_\oi\frac{R_*^2}{\ell^2}\Bigr)\sqrt{c_\ai^2+\frac{R_*^2}{\ell^2}}\;,
\end{equation}
and for the black hole asymptotics
\begin{equation}\label{eq:bubbleRadBHasymp}
    (2R_*^2-R_\hor^2)\sqrt{\ell^2-\hat\eps_\oi R_*^2} =( \ell^2-2\hat\eps_\oi R_*^2)\sqrt{R_*^2-R_\hor^2}\;.
\end{equation}
Both conditions have the same structure, as demonstrated by introducing the auxiliary variable $\xx$
\begin{equation}\label{eq:xxdef}
    \xx=
    \begin{cases}
      \ell^2c_\ai^2 &\text{for $\xx>0$\;,}\\
      -R_\hor^2   &\text{for $\xx<0$\;,}
    \end{cases}
\end{equation}
yielding to a simple quadratic equation
\begin{equation}\label{eq:xcondition}
    (\ell^2-\hat\eps_\oi R_*^2)\, \xx^2 - \Chi\, \xx - R_*^2\Chi =0\;,
\end{equation}
where
\begin{equation}
    \Chi = \ell^4 - 4R_*^2(1+\hat\eps_\oi)(\ell^2-\hat\eps_\oi R_*^2)\;.
\end{equation}
It has a solution only if
\begin{equation}\label{eq:Chipos}
    \Chi>0\;.
\end{equation}
For ${\hat\eps_\oi>-1}$, this enforces an upper limit on the size of the bubble
\begin{equation}
    R_*^2<
    \frac1{2\hat\eps_\oi}\Bigl(1-\frac1{\sqrt{1+\hat\eps_\oi}}\Bigr)
    \;.
\end{equation}
Notice that the condition ${\hat\eps_\oi>-1}$ means that the effective cosmological constant $\Lambda'$ inside the bubble is greater than the asymptotic one, ${\Lambda'>\Lambda}$. For ${\hat\eps_\oi<-1}$, i.e., $\Lambda'<\Lambda$, the condition \cref{eq:Chipos} is always satisfied.

If \cref{eq:Chipos} is satisfied, the equation \cref{eq:xcondition} has two solutions for $\xx$.
They have opposite signs, one referring to the conical asymptotics and the other to the black hole asymptotics
\begin{align}
    \label{eq:conditions}
    c_\ai^2 &= \frac{\sqrt{\Chi}}{2\ell^2(\ell^2-\hat\eps_\oi R_*^2)}\bigl(\ell^2-2\hat\eps_\oi R_*^2 + \sqrt{\Chi}\bigr)\;,\\
    R_\hor^2 &= \frac{\sqrt{\Chi}}{2(\ell^2-\hat\eps_\oi R_*^2)}\bigl(\ell^2-2\hat\eps_\oi R_*^2 - \sqrt{\Chi}\bigr)\;.
\end{align}

Next, we want to express the central energy density $\hat\eps_\oi$ and other quantities in terms of the parameters of the asymptotic solution. Let us first mention that, for the consistency of the inner and outer solutions, the individual terms under the square root in \cref{eq:bubbleRadconasymp} and \cref{eq:bubbleRadBHasymp} must be positive
\begin{equation}\label{eq:poscond1}
   \ell^2-\hat\eps_\oi R_*^2>0 \;,\quad \ell^2c_\ai^2+R_*^2>0\;,\quad R_*^2-R_\hor^2 >0\;.    
\end{equation}
Since $R_*^2>0$, conditions \cref{eq:bubbleRadconasymp}, \cref{eq:bubbleRadBHasymp} also enforce 
\begin{equation}\label{eq:poscond2}
  \ell^2-2\hat\eps_\oi R_*^2>0\;.    
\end{equation}
As we already said, both \cref{eq:bubbleRadconasymp}, \cref{eq:bubbleRadBHasymp} are identical when written using $\xx$ instead of $c_\ai$ and $R_\hor$, cf.~\cref{eq:xxdef}. They also yield a quadratic equation for $\hat\eps_\oi$ that has a real solution if 
\begin{equation}
    \Upsilon\equiv \xx^2+4(R_*^2+2\ell^2)(\xx+R_*^2)>0\;.
\end{equation}
Thanks to \eqref{eq:poscond1}, it always holds. The relevant solution is
\begin{equation}
    \hat\eps_\oi = \frac{12\ell^2(\xx+R_*^2)-\Upsilon-(\xx+2R_*^2)\sqrt{\Upsilon}}{8R_*^2(\xx+R_*^2)}\;.
\end{equation}
The other solution with $+$ sign in front of $\sqrt{\Upsilon}$ term does not satisfy the condition \cref{eq:poscond2}. 

We thus obtained a unique value of the bubble energy density $\hat\eps_\oi$ for any conical asymptotics (given by $\xx=\ell^2c_\ai^2$) and black hole asymptotics (given by $\xx=-R_\hor^2$) and a bubble size $R_*$ satisfying relevant conditions.

The local mass $\hat\locmass_-$ inside the bubble, the mass of the shell $\hat\locmass_\shl$, and the mass $\hat\locmass_+$ of the bubble, including the shell, are
\begin{align}
    \hat\locmass_- &= 1- \sqrt{\frac12\Bigl(1+\frac{R_*^2}{\ell^2}\Bigr)
        +\frac{c_\ai^4+\bigl(c_\ai^2+2\frac{R_*^2}{\ell^2}\bigr)\frac{\sqrt{\Upsilon}}{\ell^2}}{8\bigl(c_\ai^2+\frac{R_*^2}{\ell^2}\bigr)}}\;,\notag\\
    \hat\locmass_\shl &= \hat\locmass_+ - \hat\locmass_-\;,\\
    \hat\locmass_+ &= 1- \sqrt{c_\ai^2+\frac{R_*^2}{\ell^2}}\;.\notag
\end{align}
for the conical asymptotics. We get analogous expressions for the black hole asymptotics by setting ${c_\ai^2\to-\frac{R_\hor^2}{\ell^2}}$.

For the conical asymptotics, we can perform the limit $R_*\to0$ with fixed asymptotic conicity $c_\ai$. It provides the expected result
\begin{equation}
    \hat\locmass_+ \to \hat\locmass_\oi= 1- c_\ai\;.
\end{equation}

    \section{Summary}
\label{sec:summary}

We discussed a static point-like source in three-dimensional gravity, mostly in spacetime with AdS asymptotics. We modeled such a source as a matter object with a regular origin and studied the limit of negligible size of the object. 

For that, we reviewed static and rotationally symmetric solutions to the vacuum Einstein equations with a cosmological constant and discussed dust, incompressible fluid, and dark matter as candidates for the object's interior. 

In general, gluing the outer asymptotic solution to the inner spacetime with matter requires a thin shell on the joining surface. We considered a shell with a Nambu-Goto and radiation equation of state. In the case of the fluid, the gluing could be done without any shell. 

We introduced two natural notions of mass, which we employed in the discussion.

We showed that, in the limit as the matter object's size approaches zero while keeping the asymptotic characteristics of the spacetime, the geometry becomes conical, with the tip of the cone representing a point-like object. We demonstrated that the local mass of the regular object limits to the formula \cref{eq:locmassconicity}, $\locmass_\oi=1-c_\ai$. It is robust under the choice of the matter model since it depends solely on the asymptotic geometry outside the object -- as discussed at the beginning of \cref{sec:combine}. Surprisingly, it demonstrates that a natural notion of the local mass does not yield the BTZ mass parameter $\mBTZ$, which enters the famous form \cref{eq:BTZ_metric} of the static rotationally symmetric metric. As can be seen from the discussion of various matter models, neither the notion of the Killing mass is directly related to $\mBTZ$. Of course, $\mBTZ$ and $\hat\locmass_\oi$ are related by \cref{eq:locmassBTZmass}.

We thus demonstrated that the point particle in three-dimensional AdS spacetime can be well modeled as a limit of spacetimes with a sufficiently smooth mass-density regular at the origin. The conicity of the resulting spacetime is directly related to the local mass of the original object, where the local mass is just a plain integral of local energy density over a spatial time slice.

\begin{acknowledgments}
The authors thank the Czech Science Foundation grant GA\v{C}R~22-14791S. P.L. acknowledges support from the Charles University Research Center grant UNCE24/SCI/016 and the Charles University Student Science Project SVV260833.
\end{acknowledgments}
    \bibliography{8_bibliography}
    \label{sec:bib}

\end{document}